\RequirePackage{fix-cm}
\documentclass[smallextended]{svjour3}
\usepackage{amsfonts, amssymb, graphicx, float, appendix, multirow, longtable}
\usepackage{mathtools}
\usepackage{color}
\usepackage{hyperref}
\usepackage{mathrsfs}
\usepackage[left=2.5cm,top=1.5cm,right=2.5cm,bottom=3cm]{geometry}
\usepackage{subfigure}

\newcommand{\sfrac}[2]{{\textstyle{#1\over#2}}}
\newcommand{\dd}{\text{D}}
\newcommand{\tdd}{\tilde{\text{D}}}
\newcommand{\diff}{\text{d}}
\newcommand{\pp}{\partial}

\begin{document}
\fontsize{12pt}{13pt}\selectfont
\title{\textbf{On the Hyperbolicity and Stability of $3+1$ Formulations of Metric $f(R)$ Gravity.}}

\author{Bishop Mongwane}
\institute{Bishop Mongwane \at Department of Mathematics \& Applied Mathematics,
University of Cape Town, 7701 Rondebosch, South Africa \\
\email{astrobish@gmail.com}}

\maketitle

\begin{abstract}

$3+1$ formulations of the Einstein field equations have become an invaluable tool in Numerical relativity, having been used successfully in modeling spacetimes of black hole collisions, stellar collapse and other complex systems. It is plausible that similar considerations could prove fruitful for modified gravity theories. In this article, we pursue from a numerical relativistic viewpoint the $3+1$ formulation of metric $f(R)$ gravity as it arises from the fourth order equations of motion, without invoking the dynamical equivalence with Brans-Dicke theories. We present the resulting system of evolution and constraint equations for a generic function $f(R)$, subject to the usual viability conditions. We confirm that the time propagation of the $f(R)$ Hamiltonian and Momentum constraints take the same Mathematical form as in general relativity, irrespective of the $f(R)$ model. We further recast the 3+1 system in a form akin to the BSSNOK formulation of numerical relativity. Without assuming any specific model, we show that the ADM version of $f(R)$ is weakly hyperbolic and is plagued by similar zero speed modes as in the general relativity case. On the other hand the BSSNOK version is strongly hyperbolic and hence a promising formulation for numerical simulations in metric $f(R)$ theories.

\end{abstract}

\section{Introduction}
The field of Numerical relativity has seen several successes in the modeling of strong-field dynamics. For example, the once perplexing simulations of colliding black holes are now routine applications that can be executed on moderate workstations. This is a result of a journey that spans decades of interdisciplinary work, drawing from seemingly divergent fields such as geometry, numerical analysis, software design, physics etc \cite{Lehner:2001wq,0264-9381-25-9-093001,lrr-2015-1}. On the other hand, the field has reached a state of maturity that it can prove useful in other contexts such as gravity theories other than general relativity and in spacetime dimensions higher than four. Phenomenologically, these are well motivated situations as the theory of general relativity may be considered as a special case of a more general theory of gravitation \cite{Seiberg:2006wf}. With this in mind, one can gain some insights into the Cauchy problem of standard general relativity by studying the problem in a more general setting such as modified gravity.

Modified theories of gravity such as $f(R)$ and other higher order theories can account for the late time accelerated expansion of the universe without invoking an exotic dark energy component or the \textit{ad hoc} introduction of a cosmological constant. Rather, within such theories, accelerated expansion arises naturally from the concomitant field equations via curvature terms. They thus have the potential of solving one of the most important problems in 21st century cosmology. As a consequence of this, and other motivations arising from astrophysics, inflationary cosmology and high energy physics, $f(R)$ theories have been studied extensively in the literature, with applications ranging from solar system dynamics to cosmological scenarios, employing a range of tools to investigate the viability of such theories and to further constrain the parameters of the theory \cite{Sotiriou:2008rp,0004-637X-779-1-39,DeFelice:2010aj,capozziello2010beyond}. 

Because of the extra degrees of freedom inherent in the theory, studies in $f(R)$ gravity have proven interesting, often  exhibiting a number of peculiarities in comparison to similar systems in general relativity. For example, the Jebsen-Birkhoff theorem\footnote{This can be loosely stated as `Spherically symmetric vacuum solutions of the field equations are necessarily those of Schwarzschild'.} does not hold in metric $f(R)$ under general conditions, though it may be extended to Palatini $f(R)$ \cite{Sotiriou:2008rp,Nzioki:2013lca}. Other examples can be found in e.g \cite{Abebe:2011in,Abebe:2014hka,Clifton:2012ry,Ganguly:2013taa,Resco:2016upv}. In principle, it is plausible that further characteristic signatures of $f(R)$ gravity remain to be uncovered as they manifest only through non-linear terms, which are often discarded in perturbative analyses. As a result, further exploration of these and other scenarios could be hindered by the complexity of the non-linear fourth order system or the often unrealistic symmetry assumptions that are necessary if one is to obtain solutions by analytical means. In order to take advantage of the predictive power of any physical theory, however, one needs to be able to generate solutions under various configurations of interest without such limitations. Indeed, a numerical approach to $f(R)$ gravity could prove fruitful. After all, even in the simplest relativistic theory, general relativity, one needs numerical methods to model complex spacetimes such as those involving the colliding of binary black holes.

A numerical approach to fourth order gravity, or any relativistic theory for that matter, is not without problems. In numerical relativity for example, one has to decide which initial value problem to solve. In general, initial value problems are classified as either being $3+1$ Cauchy, Characteristic \cite{Bondi21,Sachs103} or Hyperboloidal \cite{hyperboloidal,lrr-2004-1}, depending  on the nature of the hypersurface on which initial data is prescribed. Even within the $3+1$ Cauchy sub-class, there are different formulations of the Einstein field equations, which although Mathematically equivalent, behave differently when discretized and evolved numerically. This is the essence of the formulation problem in numerical relativity \cite{Shinkai:2002yf,lrr-1998-3}. In general, a typical recipe for a successful numerical evolution must take into account the following points: Initial data problem \cite{lrr-2000-5}, Formulation problem \cite{Shinkai:2002yf}, Boundary problem, Gauge problem and so on, See \cite{0264-9381-25-9-093001,alcubierre2012introduction,baumgarte2010numerical,gourgoulhon20123+1} for an in-depth discussion of these and other practical aspects of numerical relativity. It is conceivable that, when considering fourth order gravity, each of these areas will be affected in non-trivial ways by the accompanying new degrees of freedom.

Under certain conditions, it can be shown that $f(R)$ gravity is dynamically equivalent to a Brans-Dicke scalar-tensor theory, with metric $f(R)$ corresponding to Brans-Dicke parameter $\omega_{0}=0$, while Palatini $f(R)$ corresponds to $\omega_{0}=-3/2$ \cite{Sotiriou:2006hs}. In fact, with the exception of \cite{Tsokaros:2013fma}, the Cauchy problem of $f(R)$ has mostly been studied by utilizing this equivalence, see \cite{PhysRevD.87.104029,PhysRevD.90.024017} for examples involving numerical applications. However, such dynamical equivalence should be interpreted with caution \cite{Jaime:2010kn,Briscese:2006xu,Capozziello:2006dj,PhysRevLett.102.221101}. The initial value problem for scalar-tensor theories was systematically presented in \cite{0264-9381-23-14-010}. This was later specialized to $f(R)$ gravity in \cite{0264-9381-24-22-024,Capozziello:2010ut,Capozziello:2011gw,Capozziello:2013gza}, largely focusing on the well-formulatedness of the problem. In \cite{2011CQGra..28h5006P}, it was shown that the Hamiltonian and Momentum constraints for $f(R)$ are preserved in time. In this work, we study the Cauchy formulation of metric $f(R)$ gravity from a numerical relativity perspective, without using the equivalence with Brans-Dicke theories. We derive the ADM \cite{Arnowitt:1962hi,1979sgrr.work...83Y} and BSSNOK \cite{Nakamura01011987,Shibata:1995we,Baumgarte:1998te} formulations for $f(R)$ gravity and show that, like in the general relativity case, the BSSNOK formulation has more appealing stability properties as a numerical formulation than the ADM version.

Although we focus only on two formulations in this work, $3+1$ formulations of the field equations are generally not limited to the ADM or BSSNOK evolution equations. There are several strategies that have been employed in the literature to construct alternative formulations, possibly using the ADM equations as a starting point \cite{Shinkai:2002yf,lrr-1998-3}. For example, one can derive new formulations by introducing additional variables based on conformal transformations of the standard ADM variables and optionally modifying the resulting evolution equations by adding arbitrary multiples of the constraints. Formulations based on this procedure have shown improved stability properties in long term evolutions. Popular examples include the BSSNOK \cite{Nakamura01011987,Shibata:1995we,Baumgarte:1998te}, Z4c \cite{Bernuzzi:2009ex,Ruiz:2010qj} and other BSSNOK-like systems \cite{Alcubierre:1999wj,Laguna:2002zc,Nagy:2004td}. In addition, one can construct fully first order hyperbolic systems by introducing spatial derivatives of existing variables as new independent variables, effectively introducing new sets of subsidiary constraints. This approach is motivated in part by the fact that first order symmetric systems are often endowed with certain desirable features such as strong hyperbolicity. Some popular examples in this category include the Bona-Masso \cite{PhysRevLett.68.1097,Bona:1997hp,PhysRevLett.75.600} and Kidder-Scheel-Teukolsky \cite{Kidder:2001tz} formulations. Another possibility is the construction of asymptotically constrained systems that seek to control the violation of constraints by having the constraint surface as an attractor. This includes the $\lambda-$system \cite{Brodbeck:1998az,Siebel:2001dr} and a family of adjusted systems \cite{PhysRevD.35.1095,Yoneda:2000zr,Yoneda:2002kg,Yoneda:2001iy}. All these have shown several advantages when compared to the standard ADM formulation \cite{Shinkai:2002yf,lrr-1998-3}.

This paper is organised as follows: In \S\ref{sec:preiminaries} we briefly introduce some of the necessary $3+1$ concepts for the purposes of fixing the context and notation. We present the field equations for fourth order gravity in \S\ref{sec:field_eqs_fr_gravity} and tackle the $3+1$ decompositions in \S\ref{sec:3p1_formulation}. We present the ADM version of the field equations in \S\ref{sec:adm_form}, and the time propagation of the constraints in \S\ref{sec:contracted_bianchi_identities}. The BSSNOK version is given in \S\ref{sec:bssnok_form}. Finally we present some analyses in \S\ref{sec:zero_speed_modes} \& \S\ref{sec:hyperbolicity} and concluding remarks in \S\ref{sec:concluding_remarks}.

\section{Preliminaries}
\label{sec:preiminaries}

In the standard $3+1$ decomposition, the metric of the spacetime is written as
\begin{equation}
ds^{2} = -(\alpha^{2} - \beta_{i} \beta^{i}) \diff t^{2} +2\beta_{i}\diff t \diff x^{i}+ \gamma_{ij} \diff x^{i} \diff x^{j}\;,
\end{equation}
where $\gamma_{ij}$ is the spatial metric induced on the hypersurface by a unit normal vector $n^{a}$, $\alpha$ is the lapse function and $\beta^{i}$ is the shift vector with $\beta_{i}=\gamma_{ij}\beta^{j}$. In terms of the normal vector $n^{a}$, the spatial metric can be written as 
\begin{equation}
\gamma_{ab} = g_{ab} + n_{a}n_{b}, \qquad \gamma^{ab} = g^{ab} + n^{a}n^{b}\;.
\end{equation}
Thus $\gamma^{a}_{\phantom{a}b}$ is a projection tensor that projects geometric objects into the spatial hypersurface, while $n^{a}n_{b}$ projects along the normal, such that,
\begin{align}
N^{a}_{\phantom{b}b} &= -n^{a}n_{b}\;,  &\Rightarrow  &\qquad N^{a}_{\phantom{a}c}N^{c}_{\phantom{a}b} = N^{a}_{\phantom{a}b}, \;  N^{a}_{\phantom{a}a}=1, \;  N_{ab}n^{b}=n_{a} \;, \\
\gamma^{a}_{\phantom{b}b} &= \delta^{a}_{\phantom{b}b} + n^{a}n_{b}\;, &\Rightarrow  &\qquad \gamma^{a}_{\phantom{a}c} \gamma^{c}_{\phantom{a}b} = \gamma^{a}_{\phantom{a}b},\; \gamma^{a}_{\phantom{a}a}=3, \;  \gamma_{ab}n^{b}=0\;.
\end{align}
 We define a spatially projected covariant derivative, compatible with the induced metric $\gamma_{ab}$, as
\begin{equation}
 \dd_{e} Q^{a\cdots b}_{\phantom{a\cdots b} c\cdots d}\equiv
\gamma^{a}_{\phantom{a}p} \cdots \gamma^{b}_{\phantom{q}q}\, \gamma^{r}_{\phantom{a}c}\cdots \gamma^{s}_{\phantom{a}d}\,
\gamma^{f}_{\phantom{e}e}\nabla_{f} Q^{p\cdots q}_{\phantom{a\cdots b} r\cdots s}\;.
\end{equation}
One can easily show that $\dd_{a}\gamma_{bc}=0$ as required.

The extrinsic curvature $K_{ab}$ can be defined in terms of projections of the first covariant derivative of $n^{a}$,
\begin{align}
K_{ab} =& -\gamma^{c}_{a}\gamma^{d}_{b}\nabla_{c}n_{d}\\
=& -\dd_{a}n_{b}-n_{a}a_{b}\;,
\end{align}
where $a_{b}=n^{c}\nabla_{c}n_{b}$ is the acceleration of the unit normal $n_{a}$, and is related to the lapse function $\alpha$ via 
\begin{equation}
a_{b}=\dd_{b}\ln\,\alpha\;.
\end{equation}
Alternatively, one can give the extrinsic curvature in terms of the metric $\gamma_{ab}$
\begin{equation}
\label{eq:kab_def}
K_{ab}=-\sfrac{1}{2}\mathcal{L}_{n}\gamma_{ab}\;,
\end{equation}
where $\mathcal{L}_{n}$ denotes the Lie derivative along $n^{a}$. Following \cite{Tsokaros:2013fma}, we also define the quantity
\begin{align}
\label{eq:psi_def}
\psi = \mathcal{L}_{n}R\;.
\end{align}
where $R$ is the Ricci scalar of the spacetime. Equations (\ref{eq:kab_def}) and (\ref{eq:psi_def}) give propagation equations for the spatial metric $\gamma_{ij}$ and the Ricci scalar $R$.

\section{Field equations for $f(R)$ gravity}
\label{sec:field_eqs_fr_gravity}
In this section, we outline the field equations for $f(R)$ gravity. We note that there are three versions of $f(R)$ modified gravity: metric formalism, Palatini formalism and metric-affine gravity \cite{Sotiriou:2006hs}. In the following, we consider exclusively the metric formalism. Following standard practice, we write the modification to the Einstein-Hilbert action as,
\begin{equation}
\label{eq:action}
S=\frac{1}{2\kappa^{2}}\int dx^{4}\left[\sqrt{-g} f(R) + 2\kappa^{2}\mathcal{L}_{m} \right]\;,
\end{equation}
where $\kappa^{2}=8\pi G$, $G$ being the standard gravitational constant, $g$ is the determinant of the space-time metric $g_{ab}$, $\mathcal{L}$ is the Lagrangian density of standard matter fields, $f(R)$ is a non-linear function of the Ricci scalar $R$ only, and it is to be understood that $f(R)=R$ corresponds to general relativity without a cosmological constant\footnote{This parametrization differs from the action considered in \cite{Tsokaros:2013fma} where the parametrization is such that general relativity is recovered for $f(R)=0$.}. Throughout this paper, we work in the Jordan Frame. 

In addition to the Ricci scalar $R$, one can also include terms which are quadratic in the Ricci tensor $R_{ab}$, Riemann tensor $R_{abcd}$ and conceivably other curvature invariants in the action (\ref{eq:action}), leading to a richer set of field equations \cite{Stelle1978,noakes:1983,PhysRevD.28.1876,PhysRevD.53.5583,Carroll:2004de,Myrzakulov:2013hca}. Typically when considering such $f(R, R_{ab}R^{ab}, R_{abcd}R^{abcd}, \cdots)$ modifications, massive gravitons, vector degrees of freedom and potentially ghost degrees of freedom associated with Ostrogradski instabilities will manifest in the resulting theory \cite{Carroll:2004de,Woodard:2006nt,Sotiriou:2008rp,DeFelice:2010aj}. The latter may be avoided by considering functions $f(R,\mathcal{G})$ of the Gauss-Bonnet invariant $\mathcal{G}=R^{2}-4R_{ab}R^{ab}+ R_{abcd}R^{abcd}$ which have found applications in Cosmology \cite{Cognola:2006eg,Nojiri:2005jg,Goheer:2009qh}. In this work, we restrict our attention to the sub-class of $f(R)$ theories arising from the action (\ref{eq:action}). Although simpler in nature, these are sufficiently general to incorporate some of the basic aspects of higher order gravity.

Varying the action (\ref{eq:action}) with respect to the metric results in the field equations
\begin{equation}
f' R_{ab}-\frac{1}{2}fg_{ab}-\nabla_{a}\nabla_{b}f' +g_{ab}\Box f'=\kappa^{2}T_{ab}\;,
\end{equation}
where $f'=\pp f(R)/\pp R$, $f=f(R)$ and $T_{ab}$ is the energy momentum tensor of standard matter fields and is defined in terms of $\mathcal{L}_{m}$ as,
\begin{equation}
T_{ab} = \frac{2}{\sqrt{-g}}\frac{\delta (\sqrt{-g}\mathcal{L}_{m})}{\delta g_{ab}}\;.
\end{equation}
Unlike in the general relativity case, $f(R)$ theories posses a massive scalar degree of freedom commonly referred to as a \textit{scalaron} \cite{STAROBINSKY198099}, whose propagation is governed by the trace of the field equations,
\begin{equation}
3\,\Box f' -2f+f'R=\kappa^{2} T\;,
\end{equation}
where $T=g^{ab}T_{ab}$. For notational simplicity, it is convenient to introduce the symmetric tensor $\Sigma_{ab}$ such that
\begin{align}
\Sigma_{ab} &= f' R_{ab}-\sfrac{1}{2}fg_{ab}-\nabla_{a}\nabla_{b}f' +g_{ab}\Box f' \\
&=  f' R_{ab}-\sfrac{1}{2}fg_{ab}-f''\nabla_{a}\nabla_{b}R - f'''\nabla_{b}R\nabla_{a}R +g_{ab}(f'''\nabla^{c}R\nabla_{c}R + f'' \Box R)\;,
\end{align}
where a repeated $'$ denotes repeated differentiation with respect to the Ricci scalar, such that
\begin{align}
\label{eq:fps}
f=f(R)\;, \qquad f'=\frac{\pp f(R)}{\pp R}\;, \qquad f''=\frac{\pp^{2} f(R)}{\pp R^{2}}\qquad \text{and} \qquad f'''=\frac{\pp^{3} f(R)}{\pp R^{3}}\;.
\end{align}
 One can therefore write the field equations compactly as,
\begin{equation}
\label{eq:compact_field_eqs}
\Sigma_{ab} = \kappa^{2} T_{ab}\;.
\end{equation}
where $\nabla^{a}\Sigma_{ab}=0=\nabla^{a}T_{ab}$ as a consequence of the Generalized Bianchi identity \cite{Koivisto:2005yk}. Alternatively, one can use the trace $\Sigma=\kappa^{2} T$ of the field equations (\ref{eq:compact_field_eqs}) to arrive at\footnote{In the general relativity case, Equation (\ref{eq:compact_field_eqs}) reduces to $\displaystyle R_{ab}-\sfrac{1}{2}g_{ab}R=\kappa^{2} T_{ab}$, while the expression (\ref{eq:compact_field_eqs_2}) reduces to the alternate form $\displaystyle R_{ab}=\kappa^{2} (T_{ab}-\sfrac{1}{2} g_{ab}T)$ for $\nu=\sfrac{1}{2}$.}
\begin{equation}
\label{eq:compact_field_eqs_2}
\Sigma_{ab} = \kappa^{2} T_{ab} + \nu \;g_{ab}(\Sigma-\kappa^{2} T)\;.
\end{equation}
where $\nu\in\mathbb{R}$ is a parameter whose choice will be made clear in the next section. The two versions of the field equations (\ref{eq:compact_field_eqs}) and (\ref{eq:compact_field_eqs_2}) are equivalent for physical systems, nevertheless, (\ref{eq:compact_field_eqs_2}) has a number of practical advantages in the context of a Cauchy formulation. We will highlight this point in Section (\ref{sec:3p1_formulation}). 

Although there is some freedom in the choice of the $f(R)$ function specifying a given model, this choice is not arbitrary. There are several non-trivial viability conditions to be satisfied for any given $f(R)$ function. For example, in order for the effective Newtonian gravitational potential ${G^{eff}=G^{Newton}/f'}$ to not change sign, one must have $f'>0$. In addition, one must have $f(R)\rightarrow R$ for $R\gg R_{0}$, where $R_{0}$ is the cosmological value of the Ricci scalar today, in order for the model to be consistent with Solar system constraints. See \cite{Sotiriou:2008rp,DeFelice:2010aj,capozziello2010beyond} for more details.

\section{The $3+1$ Formulation}
\label{sec:3p1_formulation}
The system of evolution and constraint equations is obtained by considering projections of the field equations \ref{eq:compact_field_eqs}. Considering the normal $n^{a}$ and spatial $\gamma^{a}_{\phantom{b}b}$ projectors, there can only be three distinct set of equations,
\begin{subequations}
\label{eq:projections}
\begin{align}
n^{a}n^{b}\;\Sigma_{ab} &= n^{a}n^{b} \; \kappa^{2} T_{ab} \;,\\
\gamma^{a}_{\phantom{c}c}n^{b}\; \Sigma_{ab} &= \gamma^{a}_{\phantom{c}c}n^{b} \; \kappa^{2} T_{ab} \;,\\
\label{eq:projection3}
\gamma^{a}_{\phantom{c}c} \gamma^{b}_{\phantom{d}d} \; \Sigma_{ab} &= \gamma^{a}_{\phantom{c}c} \gamma^{b}_{\phantom{d}d} \; \kappa^{2} T_{ab}\;.
\end{align}
\end{subequations}
In order to deal with the projections of the Energy momentum tensor $T_{ab}$, we introduce the fluid quantities,
\begin{equation}
\label{eq:fluid_quantities}
\rho=n^{a}n^{b}T_{ab} \;,
\qquad S_{c} = -\gamma^{a}_{\phantom{c}c}n^{b} \; T_{ab} \;
\qquad \text{and} \qquad S_{cd}=\gamma^{a}_{\phantom{c}c} \gamma^{b}_{\phantom{d}d} \;  T_{ab}
\end{equation}
where $\rho$ is the energy density of the fluid as measured by an observer $n^{a}$, $S_{a}$ is the momentum density of the matter fluid as measured by an observer $n^{a}$ and $S_{ab}$ is the spatial stress of the matter fluid, with trace $S=\gamma^{ab}S_{ab}$. The trace of the energy momentum tensor is therefore given by
\begin{equation}
T=g^{ab}T_{ab} = S-\rho\;.
\end{equation}
It is more convenient to rewrite the projections (\ref{eq:projections}) as 
\begin{subequations}
\label{eq:projections_2}
\begin{alignat}{3}
\mathcal{H} &\equiv n^{a}n^{b}\;\Sigma_{ab} -  \rho \;,\\
\mathcal{M}_{a} &\equiv -\gamma^{c}_{\phantom{a}a}n^{b}\; \Sigma_{cb} -  S_{a} \;,\\
\label{eq:projection3b}
\mathcal{E}_{cd} &\equiv \gamma^{a}_{\phantom{c}c} \gamma^{b}_{\phantom{d}d} \; \Sigma_{ab} -  S_{cd}\;,
\end{alignat}
\end{subequations}
where we have used Equations (\ref{eq:fluid_quantities}). In this case, consistency with the field equations is obtained by imposing $\mathcal{H}=0$, $\mathcal{M}_{a}=0$ and $\mathcal{E}_{ab}=0$. However, for the spatial projection, it may be more beneficial to consider instead, $\mathcal{E}_{ab} = - m_{ab}\mathcal{H}-l^{c}_{ab}\mathcal{M}_{c}$, where $m_{ab}$ and $l^{c}_{ab}$ are arbitrary tensors which are otherwise symmetric in the indices $a$ and $b$, see e.g. \cite{Frittelli:1996nj}. This parametrization is still consistent with the field equations, provided that $\mathcal{H}$ and $\mathcal{M}_{a}$ vanish. In general relativity, it is customary to restrict to the case $l^{c}_{ab}=0$ and $m_{ab}=-\mu \gamma_{ab}$, thus giving $\mathcal{E}_{ab}=\mu \gamma_{ab}\mathcal{H}$, where $\mu=0$ corresponds to the original ADM formalism while $\mu=1$ corresponds to the standard ADM formalism of York \cite{1979sgrr.work...83Y}. We note that the original ADM formalism has worse stability properties compared to the standard ADM formalism. The stability properties of the two formalisms corresponding to $\mu=0$ and $\mu=1$, respectively, was presented in \cite{Frittelli:1996nj} based on eigenvalue analyses of the propagation equations for the constraints $\mathcal{H}$ and $\mathcal{M}_{a}$. A similar constraint analysis was presented in \cite{Shinkai:2001nd} when studying adjustments of the standard ADM equations. In order to obtain the $f(R)$ equivalent of the standard ADM version, we use instead (\ref{eq:compact_field_eqs_2}) to replace the spatial projection (\ref{eq:projection3}),
\begin{align}
\label{eq:compact_field_eqs_proj}
\gamma^{a}_{\phantom{c}c} \gamma^{b}_{\phantom{d}d} \Sigma_{ab} &= \gamma^{a}_{\phantom{c}c} \gamma^{b}_{\phantom{d}d}\;\kappa^{2} T_{ab}-  \nu \gamma_{cd}(\Sigma-T)\;. 
\end{align}
It is not difficult to show that this is equivalent to (\ref{eq:projection3b}) with $\mathcal{E}_{ab}=\gamma_{ab}\mathcal{H}$ as required.
Computing the projections of $\Sigma_{ab}$ require some attention.
Utilizing the Gauss, Codazzi and Ricci geometric relations \cite{wald:1984}, and, after some tedious algebra, one can show that they reduce to
\begin{subequations}
\label{eq:lhs_projections}
\begin{align}
\label{eq:hamilton_constraint}
n^{a}n^{b}\;\Sigma_{ab} =& \sfrac{1}{2}f -\sfrac{1}{2}Rf'+ \sfrac{1}{2}(\mathcal{R}+K^{2}-K_{ab}K^{ab})f'-\dd^{a}\dd_{a}f'-K\mathcal{L}_{n}f' \\
=&\sfrac{1}{2}f -\sfrac{1}{2}Rf'+ \sfrac{1}{2}(\mathcal{R}+K^{2}-K_{ab}K^{ab})f'-(\dd^{c}\dd_{c}R+K\mathcal{L}_{n}R)f''\nonumber \\
&-f'''\dd^{c}R\dd_{c}R  \;,
\end{align}
%
\begin{align}
\label{eq:momentum_constraint}
\gamma^{a}_{\phantom{c}c}n^{b}\; \Sigma_{ab} =& -f'(\dd_{b}K_{c}^{\phantom{b}b} -\dd_{c}K)-K_{ac}\dd^{a}f'-\dd_{c}(\mathcal{L}_{n}f') \\
=&  -f'(\dd_{b}K_{c}^{\phantom{b}b} -\dd_{c}K)-f''(K_{c}^{\phantom{a}a}\dd_{a}R +\dd_{c}\mathcal{L}_{n}R) - f'''\mathcal{L}_{n}R \dd_{c}R \;,
\end{align}
\begin{align}
\label{eq:spatial_projectiont}
\gamma^{c}_{\phantom{c}a} \gamma^{d}_{\phantom{d}b} \; \Sigma_{cd} =& f'(\mathcal{R}_{ab}-2K_{a}^{\phantom{c}c}K_{bc}+KK_{ab}-\mathcal{L}_{n}K_{ab}-\sfrac{1}{\alpha}\dd_{a}\dd_{b}\alpha) -\sfrac{1}{2}f\gamma_{ab} \nonumber \\
&-[\dd_{a}\dd_{b}f'+\mathcal{L}_{n}f'K_{ab}+\gamma_{ab}(\dd^{a}\dd_{a}f'+K\mathcal{L}_{n}f'+a^{c}\dd_{c}f'-\mathcal{L}_{n}\mathcal{L}_{n}f')]\\
=&f'(\mathcal{R}_{ab}-2K_{a}^{\phantom{c}c}K_{bc}+KK_{ab}-\mathcal{L}_{n}K_{ab}-\sfrac{1}{\alpha}\dd_{a}\dd_{b}\alpha) -\sfrac{1}{2}f\gamma_{ab}\nonumber \\
& -(\dd_{a}\dd_{b}R+\mathcal{L}_{n}R K_{ab})f''-f'''\dd_{a}R\dd_{b}R+\gamma_{ab}f'''(\dd^{c}R\dd_{c}R-\mathcal{L}_{n}R\mathcal{L}_{n}R) \nonumber \\
& +\gamma_{ab}f''(\dd^{c}\dd_{c}R+K\mathcal{L}_{n}R+a^{c}\dd_{c}R-\mathcal{L}_{n}\mathcal{L}_{n}R) \;,
\end{align}
and finally the trace gives
\begin{align}
\label{eq:trace_eq}
g^{ab}\Sigma_{ab} =& -2f + Rf' + 3(\dd^{a}\dd_{a}f'+K\mathcal{L}_{n}f'+a^{c}\dd_{c}f'-\mathcal{L}_{n}\mathcal{L}_{n}f') \\
=&-2f + Rf' + 3(\dd_{c}\dd^{c}R+K\mathcal{L}_{n}R+a^{c}\dd_{c}R - \mathcal{L}_{n}\mathcal{L}_{n}R)f''\nonumber \\
&+3(\dd^{a}R\dd_{a}R-\mathcal{L}_{n}R\mathcal{L}_{n}R)f''' \;.
\end{align}
\end{subequations}

We are now in a position to comment on the advantages of using (\ref{eq:compact_field_eqs_2}) over (\ref{eq:compact_field_eqs}). The first point being that (\ref{eq:compact_field_eqs_2}) results in York's version of the ADM decomposition, which is preferred over the original ADM form in Cauchy formulations. This point is both relevant for $f(R)$ and general relativity. An added advantage, which does not arise within general relativity, is that one cannot close the system of equations by using (\ref{eq:compact_field_eqs}). This can be seen from (\ref{eq:spatial_projectiont}) which contains both time derivatives of $K_{ab}$ and $R$ (via the Lie derivative terms). One needs the trace equation in order to eliminate the latter. But what value do we need to assign to $\nu$? We note that in order to cancel out Lie derivatives of $R$ from (\ref{eq:spatial_projectiont}), one needs to add a third of the trace (\ref{eq:trace_eq}), thus giving $\nu=1/3$, see also \cite{Tsokaros:2013fma}.

\subsection{ADM formulation}
\label{sec:adm_form}
In the rest of this article, we follow common convention \cite{alcubierre2012introduction,baumgarte2010numerical,gourgoulhon20123+1} and generally refer to the standard ADM formulation of York as simply ADM, unless in cases where ambiguities may arise.
From the above information, we have that the Hamiltonian constraint (\ref{eq:hamilton_constraint}), also referred to as the Scalar constraint, is given by
\begin{align}
\label{eq:hamilton_constraint_admfr}
\mathcal{H} \equiv& f -Rf'+ (\mathcal{R}+K^{2}-K_{ab}K^{ab})f'-2(\dd^{c}\dd_{c}R+K\psi)f''\nonumber \\
&-2f'''\dd^{c}R\dd_{c}R - 2\kappa^{2} \rho = 0\;,
\end{align}
while the Momentum constraint (\ref{eq:momentum_constraint}), also known as the Vector constraint becomes
\begin{equation}
\label{eq:momentum_constraint_admfr}
\mathcal{M}_{c} \equiv \kappa^{2} S_{c}  -f'(\dd_{b}K_{c}^{\phantom{b}b} -\dd_{c}K)-f''(K_{c}^{\phantom{a}a}\dd_{a}R +\dd_{c}\psi) - f'''\psi \dd_{c}R= 0\;.
\end{equation}

Using the definitions of the extrinsic curvature $K_{ab}$ (Equation \ref{eq:kab_def}) and $\psi$ (Equation (\ref{eq:psi_def})) we arrive at evolution equations for $\gamma_{ab}$ and $R$. The time evolution of $\psi$ is obtained from the trace of the field equations, and finally the projection \ref{eq:spatial_projectiont} results in an evolution equation for $K_{ab}$. The full system of evolution equations is therefore,
\begin{subequations}
\label{eq:prop_efe}
\begin{align}
\label{eq:propR}
\pp_{t} R =& \; \alpha \psi + \beta^{i}\dd_{i}R\;, \\
\label{eq:propPsi}
\pp_{t}  \psi =&  \frac{\alpha}{3f''}\left[-2f+Rf'+3\left(\dd_{c}\dd^{c}R+K\psi+a^{c}\dd_{c}R\right)f'' \right. \\
&  \left. +3\left(\dd^{a}R\dd_{a}R-\psi^{2}\right)f''' -\kappa^{2}(S-\rho) \right] +\beta^{i}\dd_{i}\psi\\
\pp_{t}  \gamma_{ij} =&  - 2 \alpha K_{ij} +\dd_{i}\beta_{j} + \dd_{j}\beta_{i}\;, \\
\label{eq:prop_k_ij}
\pp_t K_{ij}  =& \; \alpha (\mathcal{R}_{ij}-2K_{i}^{\phantom{c}c}K_{jc}+KK_{ij}) + \frac{\alpha}{f'}\left\{\sfrac{1}{6}f\gamma_{ij}-\sfrac{1}{3}\gamma_{ij}Rf'  \right. \nonumber \\ 
&\left. -(\dd_{i}\dd_{j}R+\psi K_{ij})f''-f'''\dd_{i}R\dd_{j}R-\kappa^{2}\left[S_{ij}-\sfrac{1}{3}\gamma_{ij}(S-\rho)\right] \right\} \nonumber \\
&  -\dd_{i}\dd_{j}\alpha + \beta^{k}\pp_{k}K_{ij}+K_{kj}\pp_{i}\beta^{k} + K_{ik}\pp_{j}\beta^{k}\;.
\end{align}
\end{subequations}
The $3-$Ricci tensor $\mathcal{R}_{ij}$ in (\ref{eq:prop_k_ij}) is computed from derivatives of the metric via,
\begin{eqnarray}
\label{ricci}
\mathcal{R}_{ij} = \frac{1}{2} \gamma^{k\ell} \left( \gamma_{kj,i\ell}
+\gamma_{i\ell,kj}-\gamma_{k\ell,ij}-\gamma_{ij,k\ell}\right) 
 + \gamma^{k\ell} \left( \Gamma^m{}_{i\ell} \Gamma_{mkj} - \Gamma^m{}_{ij} \Gamma_{mk\ell} \right)\;.
\end{eqnarray}
where the $\Gamma_{abc}$ and $\Gamma^{a}_{bc}$ are Christoffel symbols of the first and second kind, respectively, computed from the $3$--metric $\gamma_{ab}$

It is trivial to show that the above system (\ref{eq:prop_efe}) reduces to the ADM form of the Einstein field equations for $f(R)=R$. The main evolution variables are $\{R,\psi, \gamma_{ij}, K_{ij}\}$, compared to only $\{\gamma_{ij},K_{ij}\}$ for general relativity. Note that there are no propagation equations for the functions $f$, $f'$, $f''$ and $f'''$. These are to be obtained analytically by using (\ref{eq:fps}), once the function $f(R)$ has been specified.
Moreover, as in the general relativity case, the field equations do not provide propagation equations for $\alpha$ and $\beta^{i}$. These are gauge functions representing the coordinate freedom of the theory. And lastly, the constraint equations (\ref{eq:hamilton_constraint_admfr}) and (\ref{eq:momentum_constraint_admfr}) do not involve the gauge variables nor time derivatives. They are elliptic partial differential equations which are usually solved only on the initial hypersurface for the purposes of generating initial data. In a typical unconstrained evolution, the constraints are not integrated in time but are used to monitor the accuracy of the numerical solution.

\subsection{Twice Contracted Bianchi identities}
\label{sec:contracted_bianchi_identities}

For any dynamical system, it is important to establish that the constraints of such a system are satisfied not only by the initial data, but also by the solution as it evolves in time. The time propagation of the Hamiltonian and Momentum constraints can be derived somewhat effortlessly by invoking the Bianchi identities as demonstrated in \cite{Frittelli:1996nj}. In $f(R)$, this takes the form
\begin{equation}
\label{generalized_biacnhi}
\nabla^{a}(\Sigma_{ab}-\kappa^{2}T_{ab}) = 0\;. 
\end{equation}
We note that in a $3+1$ setting, the field equations (\ref{eq:compact_field_eqs}) have the following induced decomposition
\begin{align}
\Sigma_{cd}-\kappa^{2} T_{cd} &= (\Sigma_{ab}-\kappa^{2} T_{ab})\delta^{a}_{\phantom{c}c}\delta^{b}_{\phantom{d}d} \nonumber \\
&= (\Sigma_{ab}-\kappa^{2} T_{ab})(\gamma^{a}_{\phantom{c}c}-n^{a}n_{c})(\gamma^{b}_{\phantom{d}d}-n^{b}n_{d}) \nonumber \\
&= \mathcal{E}_{cd}+n_{c}\mathcal{M}_{d} + n_{d}\mathcal{M}_{c}+n_{c}n_{d}\mathcal{H}\;, 
\end{align}
where we have used (\ref{eq:projections_2}). Now, the normal and spatial projections of the Bianchi identities (\ref{generalized_biacnhi}) respectively become
\begin{align}
n^{a}\nabla^{b}(\Sigma_{ab}-\kappa^{2} T_{ab}) &= n^{a}\nabla^{b}(\mathcal{E}_{ab}+n_{a}\mathcal{M}_{b} + n_{b}\mathcal{M}_{a}+n_{a}n_{b}\mathcal{H})  \\
0&= -\mathcal{E}_{ab}\dd^{b}n^{a}-\dd^{b}\mathcal{M}_{b}-2\mathcal{M}_{a}n_{b}\nabla^{b}n^{a}-\mathcal{H}\dd^{b}n_{b}-n^{b}\nabla_{b}\mathcal{H}\;,
\end{align}
and
\begin{align}
\gamma^{a}_{\phantom{c}c}\nabla^{b}(\Sigma_{ab}-\kappa^{2} T_{ab}) =& \gamma^{a}_{\phantom{c}c}\nabla^{b}(\mathcal{E}_{ab}+n_{a}\mathcal{M}_{b} + n_{b}\mathcal{M}_{a}+n_{a}n_{b}\mathcal{H})  \\
0 =& \dd^{a}\mathcal{E}_{ac}+\mathcal{E}_{bc}n^{a}\nabla_{a}n^{b}+\mathcal{M}_{c}\dd_{a}n^{a}+\mathcal{M}_{b}n_{c}n^{a}\nabla_{a}n^{b} \nonumber \\
&+\mathcal{M}_{b}\dd^{b}n_{c}+\mathcal{H}n^{b}\nabla_{b}n_{c}+n^{b}\nabla_{b}\mathcal{M}_{c}\;.
\end{align}
This results in the following propagation equation for $\mathcal{H}$
\begin{subequations}
\begin{align}
\pp_{t}\mathcal{H} =& \beta^{a}\pp_{a}\mathcal{H} + 2\alpha K \mathcal{H} - 2\alpha\gamma^{ij}\pp_{i}\mathcal{M}_{j} - 4\gamma^{ij}\pp_{j}\alpha \mathcal{M}_{i} \nonumber \\
 &+  \alpha \pp_{l}\gamma_{mk}(2\gamma^{ml}\gamma^{kj} - \gamma^{mk}\gamma^{lj}) \mathcal{M}_{j}\;.
\end{align}
while the evolution of the momentum constraint $\mathcal{M}_{i}$ is governed by
\begin{align}
\pp_{t}\mathcal{M}_{i} =&  \beta^{j}\pp_{j}\mathcal{M}_{i}  -\sfrac{1}{2}\alpha \pp_{i}\mathcal{H} -\mathcal{H}\pp_{i}\alpha + \alpha K \mathcal{M}_{i} \nonumber \\
& -\beta^{i}\gamma^{jl} \pp_{i}\gamma_{lk} \mathcal{M}_{j} + \pp_{i}\beta_{k}\gamma^{kj}\mathcal{M}_{j}\;.
\end{align}
\end{subequations}
Interestingly, this derivation, and hence the result, does not depend on the dimension $N$ of the space time, or the theory of gravity, provided the Bianchi identities hold \cite{shinkai:2004,2011CQGra..28h5006P}. We now have the desirable property that \textit{if the constraints are satisfied on some initial hypersurface, they will remain satisfied during evolution}. However, as numerical experiments have shown, one should take this statement with a grain of salt because, in practice, the constraints are only satisfied to machine precision. Therefore, depending on the properties of the formulation used, constraint violations can get amplified during numerical evolution, leading to instabilities. This quality is related to the well-posedness of the formulation, and we return to this point in \S\ref{sec:hyperbolicity}. In order to circumvent this problem, different formulations have been proposed in numerical relativity \cite{Shinkai:2002yf,lrr-1998-3}. We turn to one such formulation in the next section.

\subsection{BSSNOK formulation}
\label{sec:bssnok_form}

The BSSNOK form of the field equations, which was pioneered in a series of works \cite{Nakamura01011987,Shibata:1995we,Baumgarte:1998te}, is a conformal traceless re-formulation of the standard ADM form. This formulation has been shown to have improved numerical properties when compared to the ADM form in the case of the Einstein field equations and is perhaps one of the most popular $3+1$ formulations for numerical simulations. The starting point is the definition of new variables in terms of the ADM variables, 
\begin{subequations}
\label{eq:bssn_variables}
\begin{align}
\label{eq:bssn_phi}
\phi &= \sfrac{1}{12}\ln\;\gamma\;,\\
\label{eq:bssn_metric}
\tilde \gamma_{ij} &= e^{- 4 \phi} \gamma_{ij}\;, \\
\tilde{A}_{ij} &= e^{- 4 \phi}\left[K_{ij} - \sfrac{1}{3} \gamma_{ij} K\right]\;, \\
\tilde{\Gamma}^i &\equiv \tilde{\gamma}^{jk} \tilde{\Gamma}^{i}_{\phantom{i}jk}  \\
&= - \tilde{\gamma}^{ij}_{\phantom{ii},j}\;, 
\end{align}
\end{subequations}
where $K$ is the trace of the extrinsic curvature\footnote{The trace $K$ is evolved as an independent variable within the BSSNOK formulation, see Eq (\ref{eq:Kdot}). However, the relation $K=\gamma^{ij}K_{ij}$ may still be used in setting up initial data for $K$.} $K=\gamma^{ij}K_{ij}$, and $\gamma$ is the determinant of the physical metric $\gamma_{ij}$. Overall, one has to evolve the variables $\{R,\psi,\phi,K,\Gamma^{i},\gamma_{ij},K_{ij} \}$, a total of $19$ variables whose evolution equations will be given shortly.

In terms of the BSSNOK variables (\ref{eq:bssn_variables}), the Hamiltonian constraint (\ref{eq:hamilton_constraint}) takes the form,
\begin{align}
\mathcal{H} \equiv& \; f -Rf' + \left[e^{-4\phi}(\tilde{\mathcal{R}} -8\tdd^{i}\tdd_{i}\phi-8\tdd^{i}\phi \tdd_{i}\phi)+\sfrac{2}{3}K^{2}-\tilde{A}_{ij}\tilde{A}^{ij}\right]f'\nonumber \\
&-2(\dd^{c}\dd_{c}R+K\psi)f''-2f'''\dd^{c}R\dd_{c}R - 2\kappa^{2} \rho = 0\;,
\end{align}
while the Momentum constraint $\mathcal{M}^{a}$ is given as
\begin{align}
\mathcal{M}^{i} \equiv& \; \kappa^{2} e^{4\phi} S^{i} - \left(\tdd_{j}\tilde{A}^{ij}+6\tilde{A}^{ij}\tdd_{j}\phi-\sfrac{2}{3}\tilde{\gamma}^{ij}\dd_{j}K \right)f' \nonumber \\
&-f''\left[\left(\tilde{A}^{ij}+\sfrac{1}{3}\tilde{\gamma}^{ij}K\right)\dd_{j}R +\tilde{\gamma}^{ij}\dd_{j}\psi\right] - \tilde{\gamma}^{ij}f'''\psi \dd_{j}R= 0\;,
\end{align}
where now $\tilde{\dd}_{a}$ is the covariant derivative associated with the conformal metric $\tilde{\gamma}_{ij}$. The system of evolution equations is given by

\begin{subequations}
\begin{eqnarray}
\pp_{t} \phi & = & -\frac{1}{6}\alpha K + \beta^{k}\pp_{k}\phi + \frac{1}{6}\pp_{k}\beta^{k},\label{eq:phidot}\\
\pp_{t} \tilde{\gamma}_{ij} & = & -2\alpha\tilde{A}_{ij} + \beta^{k}\pp_{k} \tilde{\gamma}_{ij} + \tilde{\gamma}_{ik}\pp_{j}\beta^{k} + \tilde{\gamma}_{jk}\pp_{i}\beta^{k} - \frac{2}{3}\tilde{\gamma}_{ij}\pp_{k}\beta^{k},\label{eq:gdot}\\
\pp_{t} K & = &
 \frac{\alpha}{f'} \left[- \frac{1}{2}f + f''(\dd^{c}\dd_{c}R+K\psi) + f'''\dd^{c}R\dd_{c}R + \kappa^{2} \rho	\right] \nonumber \\
        && +\alpha\left(\tilde{A}_{ij}\tilde{A}^{ij}+\sfrac{1}{3}K^2\right)-\gamma^{ij}\dd_{i}\dd_{j}\alpha + \beta^{k}\pp_{k}K \;,\label{eq:Kdot}\\
\pp_{t} \tilde{A}_{ij} & = &
	\alpha\left(K\tilde{A}_{ij}-2\tilde{A}_{ik} 
   	\tilde{A}^k{}_j\right)
   	+e^{-4\phi}\big(\alpha \mathcal{R}_{ij} -\dd_{i}\dd_{j}\alpha\big)^{TF} + \nonumber \\
        && -\frac{\alpha \,e^{-4\phi}}{f'}\left\{\left[\dd_{i}\dd_{j}R+\psi e^{4\phi} (\tilde{A}_{ij}+\sfrac{1}{3}K^{}\tilde{\gamma}_{ij})\right]f'' + f'''\dd_{i}R\dd_{j}R+\kappa^{2} S_{ij}\right\}^{TF} \nonumber \\
        &&+\beta^{k}\pp_{k} \tilde{A}_{ij} + \tilde{A}_{ik}\pp_{j}\beta^{k} + \tilde{A}_{jk}\pp_{i}\beta^{k} - \frac{2}{3}\tilde{A}_{ij}\pp_{k}\beta^{k},
   	\label{eq:prop_A_ij}\\
\pp_{t}\tilde{\Gamma}^i &=& 2\alpha\left(\tilde{\Gamma}^i_{jk}
    \tilde{A}^{jk}-\frac{2}{3}\tilde{\gamma}^{ij}K_{,j}
    +6\tilde{A}^{ij}\phi_{,j}\right) 
    -2\tilde{A}^{ij}\alpha_{,j}+\tilde{\gamma}^{jk}\beta^i{}_{,jk}+
	\frac{1}{3}
    \tilde{\gamma}^{ij}\beta^k{}_{,jk}+\beta^j\tilde{\Gamma}^i{}_{,j}
    \nonumber\\
   &&-\tilde{\Gamma}^j\beta^i{}_{,j} 
    +\frac{2}{3} \tilde{\Gamma}^i\beta^j{}_{,j} - 2\alpha \kappa^{2} e^{4\phi}\frac{S^{i}}{f'} +2\alpha \frac{f''}{f'}\left[\left(\tilde{A}^{ij}+\sfrac{1}{3}\tilde{\gamma}^{ij}K\right)\dd_{j}R+\tilde{\gamma}^{ij}\dd_{j}\psi\right] \nonumber \\ 
&& +2\alpha\frac{f'''}{f'}\tilde{\gamma}^{ij}\psi\dd_{j}R .\label{eq:Gammaidot}
\end{eqnarray}
\end{subequations}

The superscript $[\cdots]^{TF}$ in (\ref{eq:prop_A_ij}) denotes the trace-free part with respect to the physical metric $\gamma_{ij}$, and for any scalar $\chi$
\begin{equation}
\dd_{i}\dd_{j}\chi = \pp_{i}\pp_{j}\chi - 4\pp_{(i}\phi\pp_{j)}\chi-\tilde{\Gamma}^{k}_{ij}\pp_{k}\chi+
2\tilde{\gamma}_{ij}\tilde{\gamma}^{kl}\pp_{k}\phi \pp_{l}\chi.
\end{equation}
The Ricci tensor $\mathcal{R}_{ij}$ is now written as a sum of two pieces
\begin{equation}
   \mathcal{R}_{ij} = \tilde{\mathcal{R}}_{ij} + \mathcal{R}^{\phi}_{ij},
\end{equation}
where,
\begin{eqnarray}
  \mathcal{R}^{\phi}_{ij}&=&-2\tilde{\dd}_i\tilde{\dd}_j\phi-2\tilde{\gamma}_{ij}
   \tilde{\dd}^k\tilde{\dd}_k\phi + 4\tilde{\dd}_i\phi\tilde{\dd}_j\phi-4\tilde{\gamma}_{ij}\tilde{\dd}^l\phi
   \tilde{\dd}_l\phi, \\
   \tilde{\mathcal{R}}_{ij}&=&-\frac{1}{2}\tilde{\gamma}^{mn}\tilde{\gamma}_{ij,mn}
     +\tilde{\gamma}_{k(i}\tilde{\Gamma}^k{}_{,j)}
     +\tilde{\Gamma}^k\tilde{\Gamma}_{(ij)k} 
     +\tilde{\gamma}^{mn}\left(2\tilde{\Gamma}^k{}_{m(i}\tilde{\Gamma}_{j)kn}
     +\tilde{\Gamma}^k{}_{in}\tilde{\Gamma}_{kmj}\right).
\end{eqnarray}
Of course, in order for the BSSNOK system to close, it must be supplemented by the propagation equations (\ref{eq:propR}) and (\ref{eq:propPsi}).
Moreover, for the BSSNOK system to be physically equivalent to the ADM system in \S \ref{sec:adm_form}, the following auxiliary constraints must be satisfied
\begin{subequations}
\begin{align}
\tilde{\Gamma}^{i} + \pp_{j}\tilde{\gamma}^{ij}&=0 \;,\\
\text{det} \;\tilde{\gamma}_{ij} -1 &= 0 \;,\\
\tilde{\gamma}^{ij}\tilde{A}_{ij} &= 0\;.
\end{align}
\end{subequations}
Finally, we note that in deriving the time evolution of the BSSNOK variables, the Hamiltonian constraint was used in (\ref{eq:Kdot}) and the Momentum constraint used in (\ref{eq:Gammaidot}). For more details on the derivations in the context of general relativity, the reader is referred to \cite{alcubierre2012introduction,baumgarte2010numerical,gourgoulhon20123+1}. 

Although the above system is certainly more complex than the ADM system presented in \S\ref{sec:adm_form} and has a larger number of variables, it possesses properties that are desirable in comparison to the ADM formulation. We look at some of these, in turn, in Sections \ref{sec:zero_speed_modes} and \ref{sec:hyperbolicity}.

\section{Zero Speed Modes}
\label{sec:zero_speed_modes}
The idea of zero speed modes was first considered in \cite{Alcubierre:1999rt}, in a quest to analytically understand the stability properties of the ADM and BSSNOK formulations. By analysing the ADM and BSSNOK formulations linearized about a flat Minkowski background, the authors of \cite{Alcubierre:1999rt} demonstrated the existence of gauge modes and constraint violating modes that ``travel'' at zero speed. These modes were conjectured to contribute to the numerical instabilities associated with the ADM formulation. The concept can be understood by considering a one-dimensional wave equation of the form
\begin{equation}
\pp_{tt}\phi-v\pp_{xx}\phi=kF(\phi,\pp_{x}\phi,\pp_{t}\phi)\;,
\end{equation}
where $F$ is some non-linear source term and $k$ is a constant. If the speed of propagation $v$ vanishes, while the forcing term on the right hand side is non-zero, $\phi$ will rapidly grow (or collapse) without bound, leading to instabilities in the evolution. 

We refer the reader to \cite{Alcubierre:1999rt} for more details on the analysis. The aim of this section is to show that the same zero speed modes that plague the ADM formulation in numerical relativity persists even in the $f(R)$ gravity context, while the BSSNOK formulation is somewhat immune from such pathologies.

\subsection{ADM formulation}
We proceed by considering linear perturbations of flat space such that
\begin{equation}
\gamma_{ij}=\delta_{ij}+h_{ij}\;,
\end{equation}
where $h_{ij}\ll 1$. Then the ADM system presented in \S\ref{sec:adm_form} takes the form
\begin{subequations}
\label{eq:fprop_adm}
\begin{align}
\label{eq:fricci}
\pp_{t} R =& \;  \psi\;, \\
\label{eq:fpsi}
\pp_{t}  \psi =&  \sfrac{1}{3f''}\left(-2f+Rf'\right)+ \dd_{c}\dd^{c}R \;,\\
\label{eq:fhij}
\pp_{t}  \gamma_{ij} =&  - 2  K_{ij}\;, \\
\label{eq:fkij}
\pp_t K_{ij}  =& \;  \mathcal{R}_{ab} + \sfrac{1}{f'}\left\{\sfrac{1}{6}f\gamma_{ab}-\sfrac{1}{3}\gamma_{ab}Rf'   -f''\dd_{c}\dd_{d}R\right\}\;,
\end{align}
\end{subequations}
where, following \cite{Alcubierre:1999rt} we have adopted the geodesic gauge, $\alpha=1$ and $\beta^{i}=0$. We next consider a Fourier decomposition of the fields such that,
\begin{align}
\psi &=\hat{\psi}\,e^{i(\omega t - k x)} \;,\\
R &= \hat{R}\, e^{i(\omega t - k x)} \;,\\
h_{ij} &= \hat{h}_{ij}\,e^{i(\omega t - k x)}\;, \\
K_{ij} &= \hat{K}_{ij} \,e^{i(\omega t - k x)}\;,
\end{align}
where, without loss of generality, we restrict the treatment along the $x$ direction. Equations (\ref{eq:fricci}) and (\ref{eq:fhij}) immediately imply
\begin{equation}
\hat{\psi}=i\omega \hat{R} \qquad \text{and}\qquad \hat{K}_{ij} = -\frac{i\omega}{2} \hat{h}_{ij}\;.
\end{equation}
Plugging these into (\ref{eq:fpsi}) and (\ref{eq:fkij}), one arrives at the system
\begin{equation}
\omega^{2}\hat{v}\simeq k^{2}M\hat{v}+S\;,
\end{equation}
where $\hat{v}$ is a vector of dimension $7$ and is given by
\begin{equation}
\hat{v} = [\hat{R}, \hat{h}_{xx},\hat{h}_{yy},\hat{h}_{zz},\hat{h}_{xy},\hat{h}_{xz},\hat{h}_{yz}]\;.
\end{equation}
The quantity $S$ is a vector of source terms containing non-principal parts and does not affect the characteristic structure of the system \cite{gustafsson1995time}. And lastly, the $7$-by-$7$ matrix $M$ is given by
\[
M=
\left(
\begin{array}{ccccccc}
1 & 0& 0& 0& 0& 0& 0   \\
2f''/f' & 0& 1& 1& 0& 0& 0   \\
0 & 0& 1& 0& 0& 0& 0   \\
0 & 0& 0& 1& 0& 0& 0   \\
0 & 0& 0& 0& 0& 0& 0   \\
0 & 0& 0& 0& 0& 0& 0   \\
0 & 0& 0& 0& 0& 0& 1   
\end{array}
\right)\;.
\]
The eigenvalues of the matrix $M$ are given by
\begin{equation}
\lambda = 0 \;\;(\text {with multiplicity $3$}) \qquad \text{and} \qquad \lambda=1 \;\;(\text{with multiplicity $4$})\;.
\end{equation}
The $\lambda=0$ modes are the same as those in general relativity \cite{Alcubierre:1999rt}. Interestingly, these eigenvalues are independent of the function $f(R)$ or its derivatives.

\subsection{BSSNOK formulation}
We again consider linear perturbations of flat space such that 
\begin{equation}
\tilde{\gamma}_{ij}=\delta_{ij}+\tilde{h}_{ij}\;, \qquad \tilde{h}_{ij}\ll 1\;.
\end{equation}
Then the BSSNOK system presented in \S\ref{sec:bssnok_form} takes the form
\begin{subequations}
\label{eq:f2system}
\begin{align}
\label{eq:f2ricci}
\pp_{t} R =& \;  \psi\;,\\
\label{eq:f2psi}
\pp_{t}  \psi =&  \sfrac{1}{3f''}\left(-2f+Rf'\right)+ \dd_{c}\dd^{c}R \;,\\
\label{eq:f2phi}
\pp_{t} \phi =& -\sfrac{1}{6} K \;,\\
\label{eq:f2hij}
\pp_{t} \tilde{h}_{ij} =& -2\tilde{A}_{ij}\;,\\
\label{eq:f2k}
\pp_{t} K =& \sfrac{1}{f'} \left(- \sfrac{1}{2}f + f''\dd^{c}\dd_{c}R\right) \;,\\
\pp_{t} \tilde{A}_{ij} =&\big(\mathcal{R}_{ij} -\sfrac{1}{f'}\dd_{i}\dd_{j}R \big)^{TF}\;,\\
\pp_{t}\tilde{\Gamma}^i =& -\sfrac{4}{3}\tilde{\gamma}^{ij}K_{,j}-2\tilde{A}^{ij}_{\phantom{ij},j} +2 \frac{f''}{f'}\tilde{\gamma}^{ij}\dd_{j}\psi\;,
\end{align}
\end{subequations}
where we have again adopted the geodesic gauge ($\alpha=1$, $\beta^{i}=0$). We proceed by considering a Fourier decomposition of the fields so that we have plane wave solutions of the form
\begin{align}
\phi &= \hat{\phi}\,e^{i(\omega t - k x)}\;,  & K &= \hat{K}\,e^{i(\omega t - k x)} \;,\\
\psi &=\hat{\psi}\,e^{i(\omega t - k x)}\;,  & \tilde{h}_{ij} &= \hat{\tilde{h}}_{ij}\,e^{i(\omega t - k x)} \;,\\
R &= \hat{R}\, e^{i(\omega t - k x)}\;,  & \tilde{A}_{ij} &= \hat{\tilde{A}}_{ij} \,e^{i(\omega t - k x)}\;,
\end{align}
propagating along the $x$ direction.
Equations (\ref{eq:f2ricci}), (\ref{eq:f2phi}) and (\ref{eq:f2hij}) immediately imply
\begin{align}
\hat{\psi}=i\omega \hat{R}\;, \qquad \hat{K}=-6i\omega \hat{\phi}\;, \qquad \text{and} \qquad \hat{A}_{ij}=-\frac{i\omega}{2}\hat{h}_{ij}\;.
\end{align}
We can then rewrite the system (\ref{eq:f2system}) as
\begin{equation}
\omega^{2}\hat{v}\simeq k^{2}M\hat{v}+S\;,
\end{equation}
where $\hat{v}$ is a vector of dimension $8$, given by
\begin{equation}
\hat{v} = [\hat{\phi},\hat{R}, \hat{h}_{xx},\hat{h}_{yy},\hat{h}_{zz},\hat{h}_{xy},\hat{h}_{xz},\hat{h}_{yz}]\;,
\end{equation}
and again $S$ is a vector of source terms containing the non-principal parts and does not play a role in the characteristic structure of the system \cite{gustafsson1995time}. The $8$-by-$8$ matrix $M$ is now given by
\[
M=
\left(
\begin{array}{cccccccc}
0 & -\frac{f''}{6f'} & 0& 0& 0& 0& 0 &0   \\
0 & 1 & 0& 0& 0& 0& 0& 0   \\
-8 & -6\frac{f''}{f'} & 1& 0& 0& 0& 0& 0   \\
4 & 0 & 0& 1& 0& 0& 0& 0   \\
4 & 0 & 0& 0& 1& 0& 0& 0   \\
0 & 0& 0& 0& 0& 1& 0 & 0 \\
0 & 0& 0& 0& 0& 0& 1 & 0 \\
0 & 0& 0& 0& 0& 0& 0 & 1 
\end{array}
\right)\;.
\]
The eigenvalues of $M$ are given by,
\begin{equation}
\lambda = 0 \;\;(\text {with multiplicity $1$}) \qquad \text{and} \qquad \lambda=1 \;\;(\text{with multiplicity $7$})\;.
\end{equation}
Again, there is no dependence on the function $f(R)$ or its derivatives.

\section{Hyperbolicity}
\label{sec:hyperbolicity}
One of the most important considerations when dealing with numerical formulations is the concept of Hyperbolicity \cite{Gundlach:2005ta,Nagy:2004td,PhysRevD.70.104004,PhysRevD.70.044032,PhysRevD.66.064002}. Consider a PDE system written in full first order form
\begin{equation}
\label{eq:class_hyperbolic}
\pp_{t}u + M^{i}\nabla_{i}u=S(u)\;,
\end{equation}
where $u$ is a solution vector with $n$ components denoting the fundamental variables, each matrix $M^{i}$ is an $n$-by-$n$ matrix called a characteristic matrix, $i$ runs over the spatial dimensions and $S(u)$ is a source vector that may depend on the fundamental variables $u$ but not on their derivatives. The principal symbol of the system is defined by $P=M^{i}n_{i}$ where $n_{i}$ is an arbitrary unit vector. Then the system (\ref{eq:class_hyperbolic}) is said to be: (1) Weakly hyperbolic if the matrix $P$ has real eigenvalues, but does not posses a complete set of eigenvectors, (2) Strongly hyperbolic if the eigenvalues of the matrix $P$ are real, and in addition $P$ is diagonalizable for all unit vectors $n_{i}$, i.e. P has a complete set of eigenvectors, (3) Symmetric hyperbolic if the matrix $P$ is Hermitian, and (4) Strictly hyperbolic if the eigenvalues of $P$ are real and distinct. All symmetric hyperbolic systems are, by extension, strongly hyperbolic \cite{gustafsson1995time}.

This classification relates to the well posedness of a system in that, strongly hyperbolic systems are well-posed, while weakly hyperbolic systems are ill-posed. To illustrate this point, we note that for strongly Hyperbolic systems, one can always find a positive definite Hermitian matrix $H(n_{i})$, referred to as the symmetrizer such that
\begin{equation}
\label{eq:symetrizer}
HP-P^{T}H^{T} = HP - P^{T}H=0\;,
\end{equation}
where $[\cdots]^{T}$ denotes a transpose operation. One can then use $H$ to construct the energy norm for the solutions of (\ref{eq:class_hyperbolic})
\begin{equation}
\label{eq:norm}
||u||^{2} = \langle u, u \rangle = u^{\dagger}Hu\;,
\end{equation}
where $u^{\dagger}$ is the adjunct of $u$. Next, consider a Fourier mode of the form $u(x,t)=\hat{u}(t)e^{ikxn}$. Together with the evolution equation (\ref{eq:class_hyperbolic}), we can estimate the growth in the energy norm over time, by
\begin{align}
\pp_{t}||u||^{2}&=\pp_{t}(u^{\dagger})Hu + u^{\dagger}H\pp_{t}(u) \\
&= ik\hat{u}^{T}P^{T}H\hat{u} - ik\hat{u}^{T}HP\hat{u}\\
&= ik\hat{u}^{T}(P^{T}H-HP)\hat{u} \\
&=0
\end{align}
where we have used (\ref{eq:symetrizer}) and (\ref{eq:norm}). Clearly, there is no growth in the energy norm for evolution systems for which (\ref{eq:symetrizer}) holds. One therefore expects strongly hyperbolic formulations of the field equations to conserve the constraint equations during numerical evolution.

In what follows, we show that the ADM formulation (\S\ref{sec:adm_form}) is weakly hyperbolic while the BSSNOK formulation (\ref{sec:bssnok_form}) is strongly hyperbolic by studying the characteristic structures of the systems. Instead of explicitly computing the principal symbol $P$, we will proceed by adopting the more elegant method outlined in \cite{alcubierre2012introduction}. For brevity, we introduce the shorthand notation $\pp_{0}=(\pp_{t}-\mathcal{L}_{\beta})$. We will further assume that the shift vector $\beta^{i}$ is a known function of space and time, and that the lapse is a dynamical quantity whose evolution is given by a slicing condition of the Bona-Masso family\footnote{Note that the function $\zeta(\alpha)$ is customarily denoted as $f(\alpha)$ in the literature. However, we use $\zeta(\alpha)$ in this work to avoid possible confusion with the $f(R)$ function.} \cite{PhysRevLett.75.600}
\begin{equation}
\label{eq:bona_masso_slicing}
\pp_{0}\alpha = -\alpha^{2}\zeta(\alpha) K\;,
\end{equation}
where $K$ is the trace of the extrinsic curvature.

\subsection{ADM formulation}
The ADM formulation as given in \S\ref{sec:adm_form} is first order in time and second order in space. It is thus not possible to apply the above definition of Hyperbolicity to this system as it is not of the form given by Equation (\ref{eq:class_hyperbolic}). As a result, Hyperbolicity analysis of the ADM system is typically done on a first order reformulated version, similar to the Kidder-Scheel-Teukolsky formulation \cite{Kidder:2001tz,alcubierre2012introduction}. In order to proceed with the analysis, we re-cast the ADM system into fully first order form, by introducing the following quantities
\begin{equation}
r_{i} = \pp_{i}R \;, \qquad a_{i} = \pp_{i} \ln\,\alpha \;, \qquad d_{ijk}=\frac{1}{2}\pp_{i}\gamma_{jk}\;,
\end{equation}
such that the main variables of the first order system are $u=(\psi,r_{i}, a_{i},d_{ijk}, K_{ij})$, a total of $31$ quantities. The system, up to principal part, is now
\begin{subequations}
\begin{align}
\label{eq:capsi}
\pp_{0}\, \psi \simeq& \;\alpha \gamma^{ij}\pp_{i}r_{j}\;,\\
\label{eq:car}
\pp_{0}\, r_{i} \simeq&\; \alpha \pp_{i}\psi \;,\\
\pp_{0}\, a_{i}\simeq& -\alpha \zeta(\alpha)\pp_{i}K\;, \\
\pp_{0}\, d_{ijk} \simeq& -\alpha \pp_{i}K_{jk}\;,\\
\pp_{0}\, K_{ij} \simeq& -\alpha \pp_{m}\lambda^{m}_{\phantom{m}ij}\;,
\end{align}
\end{subequations}
where the quantity $\lambda^{a}_{\phantom{a}bc}$ is given by
\begin{align}
\lambda^{m}_{\phantom{m}ij} = d^{m}_{\phantom{ij}ij} + \delta^{m}_{\phantom{k}(i}\left(a_{j)} + \frac{f''}{f'}r_{j)}+ d_{j)k}^{\phantom{j)k}k} - 2d^{k}_{\phantom{k}kj)}  \right)\;.
\end{align}
If this first order reduced ADM system is strongly hyperbolic, then we should be able to find $31$ independent eigenfields, indicating the existence of a complete set of eigenvectors of the principal symbol. However, as we show below, this is not possible. If we consider derivatives only along the $x$ direction, we immediately find a set of $16$ eigenfields propagating with speed $\lambda=-\beta^{x}$,
\begin{equation}
r_{q}, a_{q}, d_{qij} \qquad q\neq x\;.
\end{equation}
Next, by considering the evolution of $\lambda^{x}_{\phantom{m}pq} (p,q\neq x)$, which can be constructed from the above system as,
\begin{align}
\pp_{0} \,\lambda^{x}_{\phantom{x}pq} =& \;\pp_{0}\, d^{x}_{\phantom{x}pq} \\
\simeq& -\alpha \gamma^{xx}\pp_{x}K_{pq}\;,
\end{align}
we find $6$ more eigenfields
\begin{equation}
\sqrt{\gamma^{xx}}K_{pq}\mp\lambda^{x}_{\phantom{x}pq}\;,
\end{equation}
traveling with speed 
\begin{equation}
\lambda_{\pm} = -\beta^{x}\pm \alpha \sqrt{\gamma^{xx}}\;.
\end{equation}
We now have $22$ out of the possible $31$ eigenfields. The two scalar equations (\ref{eq:capsi}) and (\ref{eq:car}) result in $2$ more eigenfuncitons
\begin{equation}
r^{x} \mp\sqrt{\gamma^{xx}}\psi\;,
\end{equation}
traveling with speeds 
\begin{equation}
\lambda_{\pm} = -\beta^{x}\pm\alpha\sqrt{\gamma^{xx}}\;.
\end{equation}
By considering the time evolutions of $\lambda^{x}_{\phantom{x}xq}$, given by
\begin{align}
\pp_{0}\lambda^{x}_{\phantom{x}xq} = \alpha \gamma^{xp}\pp_{x}K_{pq}\;,
\end{align}
and that of $K_{xq}$, one can notice that although the evolution of $K_{xq}$ depends on derivatives of $\lambda^{x}_{\phantom{x}xq}$, $\pp_{0}\lambda^{x}_{\phantom{x}xq}$ is essentially independent of $K_{xp}$. Therefore this subsystem cannot be diagonalized, i.e it is not possible to find a complete set of eigenfields for the ADM system when written in first order form.

\subsection{BSSNOK formulation}

Like in the previous section, we introduce a set of quantities in order to rewrite the BSSNOK system (\S\ref{sec:bssnok_form}) into fully first order form,
\begin{equation}
r_{i} = \pp_{i}R \;, \qquad a_{i} = \pp_{i} \ln\,\alpha \;, \qquad \tilde{d}_{ijk}=\frac{1}{2}\pp_{i}\tilde{\gamma}_{jk}\;, \qquad \Phi_{i}=\pp_{i}\phi\;.
\end{equation}
Then we consider the $34$ quantities\footnote{Not $38$, because $\tilde{A}_{ij}$ is traceless and $\tilde{\gamma}^{jk}\tilde{d}_{ijk}=0$ .} $u=(\psi,r_{i},a_{i}, \Phi_{i}, \tilde{d}_{ijk}, K, \tilde{A}_{ij}, \tilde{\Gamma}^{i})$ whose evolution, up to principal part is,
\begin{subequations}
\begin{align}
\label{eq:cbpsi}
\pp_{0}\, \psi \simeq& \;\alpha e^{-4\phi}\tilde{\gamma}^{ij}\pp_{i}r_{j}\;,\\
\label{eq:cbr}
\pp_{0}\, r_{i} \simeq&\; \alpha \pp_{i}\psi\;,\\
\label{eq:cba}
\pp_{0}\, a_{i}\simeq& -\alpha \zeta(\alpha)\pp_{i}K\;, \\
\pp_{0}\, \Phi_{i} \simeq& -\frac{1}{6}\alpha \pp_{i} K \;,\\
\pp_{0}\, \tilde{d}_{ijk} \simeq& -\alpha \pp_{i}\tilde{A}_{jk}\;,\\
\label{eq:cbk}
\pp_{0}\, K \simeq& -\alpha e^{-4\phi}\tilde{\gamma}^{ij}\left(\pp_{i}a_{j}-\frac{f''}{f'}\pp_{i}r_{j}\right)\;,\\ 
\label{eq:cbaa}
\pp_{0}\, \tilde{A}_{ij} \simeq& -\alpha e^{-4\phi} \pp_{k}\tilde{\lambda}^{k}_{\phantom{m}ij} \;,\\
\pp_{0}\, \Gamma^{i} \simeq& -2\alpha \tilde{\gamma}^{ij}\left(\frac{2}{3}\pp_{j}K -\frac{f''}{f'}\pp_{j}\psi \right)\;,
\end{align}
\end{subequations}
where we have introduced the quantity $\tilde{\lambda}^{m}_{\phantom{m}jk}$,
\begin{equation}
\tilde{\lambda}^{m}_{\phantom{m}ij} = \left[\tilde{d}^{m}_{\phantom{m}ij} + \delta^{m}_{\phantom{m}(i} \left(a_{j)} + \frac{f''}{f'}r_{j)}-\tilde{\Gamma}_{j)} +2\Phi_{j)} \right) \right]^{TF}\;,
\end{equation}
whose evolution is given by
\begin{align}
\label{eq:cbl}
\pp_{0} \tilde{\lambda}^{m}_{\phantom{m}ij} = -\alpha \left\{\tilde{\gamma}^{ml}\pp_{l}\tilde{A}_{ij}+\delta^{m}_{\phantom{m}(i}\left[(\zeta(\alpha)-1)\pp_{j)}K+\frac{f''}{f'}\pp_{j)}\psi \right]\right\}^{TF}\;.
\end{align}
If the BSSNOK formulation is strongly hyperbolic, then we should be able to find $34$ independent eigenfields, indicating the existence of a complete set of eigenvectors of the principal symbol. As we show below, this is indeed the case.

Considering again only derivatives along the $x$ direction, we immediately recover $20$ eigenfields propagating with speed $\lambda=-\beta^{x}$,
\begin{equation}
r_{q},a_{q},\Phi_{q},\tilde{d}_{qij},a_{x}-6\zeta(\alpha)\Phi_{x},\tilde{\Gamma}^{i}-8\tilde{\gamma}^{ij}\Phi_{j}-2\frac{f''}{f'}\tilde{\gamma}^{ij}r_{j} \qquad q\neq x\;.
\end{equation}
Next, by considering Equations (\ref{eq:cbpsi}) and (\ref{eq:cbr}), we recover $2$ more eigenfields
\begin{equation}
r^{x} \mp e^{-2\phi}\sqrt{\tilde{\gamma}^{xx}}\psi
\end{equation}
traveling with speeds
\begin{equation}
\lambda_{\pm} = -\beta^{x}\pm e^{-2\phi}\alpha\sqrt{\tilde{\gamma}^{xx}}\;.
\end{equation}
By combining Equations (\ref{eq:cbpsi}), (\ref{eq:cbr}), (\ref{eq:cba}) and (\ref{eq:cbk}) we find $2$ more eigenfields
\begin{equation}
\label{eq:oo_eig}
\left[a^{x}+\left(\frac{\zeta(\alpha)}{1-\zeta(\alpha)}\right)\frac{f''}{f'}r^{x} \right] \mp \alpha e^{-2\phi}\sqrt{\zeta(\alpha)\tilde{\gamma}^{xx}}\left[K-\left(\frac{\zeta(\alpha)}{1-\zeta(\alpha)}\right)\frac{f''}{f'}\psi \right]
\end{equation}
which travel at speeds
\begin{equation}
\lambda_{\pm} = -\beta^{x} \pm e^{-2\phi}\sqrt{\zeta(\alpha)\tilde{\gamma}^{xx}}\;.
\end{equation}
We now need $10$ more eigenfields. Equations (\ref{eq:cbaa}) and (\ref{eq:cbl}) lead to the $4$ eigenfields
\begin{equation}
\tilde{\lambda}^{xx}_{\phantom{xx}q}\mp e^{2\phi}\sqrt{\tilde{\gamma}^{xx}}\tilde{A}^{x}_{\phantom{x}q}
\end{equation}
with speeds
\begin{equation}
\lambda_{\pm}=\beta^{x}\pm \alpha e^{-2\phi}\sqrt{\tilde{\gamma}^{xx}}\;.
\end{equation}
An additional $4$ eigenfields with speeds
\begin{equation}
\lambda_{\pm}=-\beta^{x}\pm\alpha e^{-2\phi}\sqrt{\tilde{\gamma}^{xx}}
\end{equation}
are given by
\begin{equation}
e^{2\phi}\sqrt{\tilde{\gamma}^{xx}}\left(\tilde{A}_{pq}+\frac{\tilde{\gamma}_{pq}}{2\tilde{\gamma}^{xx}} \tilde{A}^{xx}\right)\mp \left(\tilde{\lambda}^{x}_{\phantom{x}pq}+\frac{\tilde{\gamma}_{pq}}{2\tilde{\gamma}^{xx}}\tilde{\lambda}^{xxx} \right)\;.
\end{equation}
And finally, the remaining $2$ eigenfields are
\begin{equation}
\label{eq:ao_eig}
  \tilde{\lambda}^{xxx}-\frac{2}{3}\tilde{\gamma}^{xx}\tilde{a}^{x}-\frac{1}{3}\tilde{\gamma}^{xx}\frac{f''}{f'}\tilde{r}^{x} \mp e^{2\phi}\sqrt{\tilde{\gamma}^{xx}}\left(\tilde{A}^{xx} -\frac{2}{3}\tilde{\gamma}^{xx}K + \tilde{\gamma}^{xx}\frac{f''}{f'}\tilde{r}^{x}\right)
\end{equation}
and they have speeds
\begin{equation}
\lambda_{\pm} = -\beta^{x}+\alpha e^{-2\phi}\sqrt{\tilde{\gamma}^{xx}}\;.
\end{equation}
In (\ref{eq:ao_eig}), we have defined $\tilde{a}^{i}=\tilde{\gamma}^{ij}a_{j}$ and $\tilde{r}^{i}=\tilde{\gamma}^{ij}r_{j}$.
An interesting thing happens when one chooses the Harmonic slicing, corresponding to $\zeta(\alpha)=1$ in (\ref{eq:bona_masso_slicing}): the eigenfields (\ref{eq:oo_eig}) become singular. However, this is not catastrophic as this case can be easily avoided by not choosing the Harmonic slicing, or by modifying the slicing condition as was done for example in the scalar-tensor case in \cite{Salgado:2008xh}.

\section{Concluding Remarks}
\label{sec:concluding_remarks}

In this paper, we have constructed and analyzed two first order in time and second order in space formulations of metric $f(R)$ in the Jordan frame representation. These are based on the standard $3+1$ ADM formulation of York \cite{Arnowitt:1962hi,1979sgrr.work...83Y} and the BSSNOK formulation \cite{Nakamura01011987,Shibata:1995we,Baumgarte:1998te}. We find that, like in general relativity, the ADM version of metric $f(R)$ is weakly hyperbolic, while the BSSNOK formulation is strongly hyperbolic and, accordingly, well-posed. In addition, the BSSNOK formulation of metric $f(R)$ has, like in general relativity, no constraint violating zero speed modes \cite{Alcubierre:1999rt}. This is a reassuring outcome as one would ordinarily expect $f(R)$ modifications to alter the characteristic structure of the $3+1$ problem. Some key features of the work presented here is that (i) we work in the Jordan frame, which is associated with the `physical' frame, and (ii) we do not utilize the dynamical equivalence between metric $f(R)$ and Brans-Dicke theories. As pointed out in \cite{Tsokaros:2013fma}, this equivalence can and has sometimes led to confusion about the true degrees of freedom for $f(R)$. Moreover, for a given $f(R)$ model, transforming to an equivalent scalar-tensor theory can lead to a multivalued scalar field potential \cite{Hindawi:1995cu}. See, for example, \cite{Jaime:2010kn} and references therein for more details on this issue. 

Within the wider context of the formulation problem, the analyses presented in \S\ref{sec:zero_speed_modes} and \S\ref{sec:hyperbolicity} are by no means exhaustive. There are other methods of analysing numerical formulations. For example in the general relativity case, eigenvalue analysis of the constraint evolution equations has been used to shed some light on the stability properties of some formulations \cite{Frittelli:1996nj,Shinkai:2001nd}. In addition, the methods presented in \cite{Calabrese:2005ft} also take into account the effect of discretization of first order in time-second order in space formulations by finite difference methods. Ultimately, it is desirable to supplement studies of numerical formulations by numerical experiments, preferably using test cases with known analytical solutions such as those presented in \cite{Alcubierre:2003pc,Babiuc:2007vr}. 

This work forms part of a larger objective aimed at constructing fully dynamical numerical simulations in metric $f(R)$ gravity using the numerical code described in \cite{bish_thesis:2014,mongwane:2015}. With the recent detection of gravitational waves by the LIGO and Virgo collaborations \cite{PhysRevLett.116.061102,Abbott:2016nmj}, it may be time to consider simulations of binary black hole collisions in $f(R)$ and other modified gravity theories with a view of exploring characteristic signatures of modified gravity in the non-linear regime. However, several issues still need to be addressed before that venture can take off. In particular, the construction of realistic binary black hole initial data for $f(R)$, the development of singularity avoiding slicings and other stable gauge conditions in general as we expect that $f(R)$ corrections will affect the well known gauge conditions that have worked so well for general relativity. 

Finally, we note that the field equations for metric $f(R)$ in the Jordan frame are fourth order. By comparison, the field equations of general relativity are only second order. In addition, the theory has more degrees of freedom and thus more generality (and complexity) than the Einstein field equations. This presents a natural arena to solidify and further advance some of the underpinning concepts of numerical relativity. For example, in the context of the formulation problem, this may be used to provide further insight into why some strongly hyperbolic formulations of the Einstein field equations seem to work better than others. Moreover, it would be interesting to study the efficacy of current gravitational wave extraction methods in capturing the scalar mode. Further study on the dynamics of the scalar mode in non linear contexts and the associated signatures in waveform extraction is underway.

\section{Acknowledgments}
The author is grateful to Obinna Umeh for many fruitful discussions and comments.







\begin{thebibliography}{10}

\bibitem{Abbott:2016nmj}
B.~P. Abbott et~al.
\newblock {GW151226: Observation of Gravitational Waves from a 22-Solar-Mass
  Binary Black Hole Coalescence}.
\newblock {\em Phys. Rev. Lett.}, 116(24):241103, 2016 arXiv:1606.04855,
  [gr-qc].

\bibitem{PhysRevLett.116.061102}
B.~P. Abbott et~al.
\newblock Observation of gravitational waves from a binary black hole merger.
\newblock {\em Phys. Rev. Lett.}, 116:061102, Feb 2016.

\bibitem{Abebe:2014hka}
A.~Abebe.
\newblock {Anti-Newtonian cosmologies in f(R) gravity}.
\newblock {\em Class. Quant. Grav.}, 31:115011, 2014 arXiv:1401.3596,  [gr-qc].

\bibitem{Abebe:2011in}
A.~Abebe, R.~Goswami, and P.~K.~S. Dunsby.
\newblock {On Shear-free perturbations of f(R) gravity}.
\newblock {\em Phys. Rev.}, D84:124027, 2011 arXiv:1108.2900,  [gr-qc].

\bibitem{alcubierre2012introduction}
M.~Alcubierre.
\newblock {\em Introduction to 3+1 Numerical Relativity}.
\newblock International Series of Monographs on Physics. OUP Oxford, 2012.

\bibitem{Alcubierre:2003pc}
M.~Alcubierre, G.~Allen, C.~Bona, D.~Fiske, T.~Goodale, et~al.
\newblock {Toward standard testbeds for numerical relativity}.
\newblock {\em Class.Quant.Grav.}, 21:589, 2004 arXiv:gr-qc/0305023,  [gr-qc].

\bibitem{Alcubierre:1999rt}
M.~Alcubierre, G.~Allen, B.~Bruegmann, E.~Seidel, and W.-M. Suen.
\newblock {Towards an understanding of the stability properties of the (3+1)
  evolution equations in General Relativity}.
\newblock {\em Phys. Rev.}, D62:124011, 2000 arXiv:gr-qc/9908079,  [gr-qc].

\bibitem{Alcubierre:1999wj}
M.~Alcubierre, B.~Bruegmann, M.~A. Miller, and W.-M. Suen.
\newblock {A Conformal hyperbolic formulation of the Einstein equations}.
\newblock {\em Phys. Rev.}, D60:064017, 1999 arXiv:gr-qc/9903030,  [gr-qc].

\bibitem{Arnowitt:1962hi}
R.~L. Arnowitt, S.~Deser, and C.~W. Misner.
\newblock {The Dynamics of general relativity}.
\newblock {\em Gen. Rel. Grav.}, 40:1997--2027, 2008 arXiv:gr-qc/0405109,
  [gr-qc].

\bibitem{Babiuc:2007vr}
M.~Babiuc, S.~Husa, D.~Alic, I.~Hinder, C.~Lechner, et~al.
\newblock {Implementation of standard testbeds for numerical relativity}.
\newblock {\em Class.Quant.Grav.}, 25:125012, 2008 arXiv:0709.3559,  [gr-qc].

\bibitem{PhysRevD.28.1876}
N.~H. Barth and S.~M. Christensen.
\newblock Quantizing fourth-order gravity theories: The functional integral.
\newblock {\em Phys. Rev. D}, 28:1876--1893, Oct 1983.

\bibitem{Baumgarte:1998te}
T.~W. Baumgarte and S.~L. Shapiro.
\newblock {On the numerical integration of Einstein's field equations}.
\newblock {\em Phys.Rev.}, D59:024007, 1999 arXiv:gr-qc/9810065,  [gr-qc].

\bibitem{baumgarte2010numerical}
T.~Baumgarte and S.~Shapiro.
\newblock {\em Numerical Relativity: Solving Einstein's Equations on the
  Computer}.
\newblock Cambridge University Press, 2010.

\bibitem{Bernuzzi:2009ex}
S.~Bernuzzi and D.~Hilditch.
\newblock {Constraint violation in free evolution schemes: Comparing BSSNOK
  with a conformal decomposition of Z4}.
\newblock {\em Phys. Rev.}, D81:084003, 2010 arXiv:0912.2920,  [gr-qc].

\bibitem{PhysRevD.70.104004}
H.~Beyer and O.~Sarbach.
\newblock Well-posedness of the baumgarte-shapiro-shibata-nakamura formulation
  of einstein's field equations.
\newblock {\em Phys. Rev. D}, 70:104004, Nov 2004.

\bibitem{PhysRevLett.68.1097}
C.~Bona and J.~Mass\'o.
\newblock Hyperbolic evolution system for numerical relativity.
\newblock {\em Phys. Rev. Lett.}, 68:1097--1099, Feb 1992.

\bibitem{Bona:1997hp}
C.~Bona, J.~Masso, E.~Seidel, and J.~Stela.
\newblock {First order hyperbolic formalism for numerical relativity}.
\newblock {\em Phys. Rev.}, D56:3405--3415, 1997 arXiv:gr-qc/9709016,  [gr-qc].

\bibitem{PhysRevLett.75.600}
C.~Bona, J.~Mass\'o, E.~Seidel, and J.~Stela.
\newblock New formalism for numerical relativity.
\newblock {\em Phys. Rev. Lett.}, 75:600--603, Jul 1995.

\bibitem{Bondi21}
H.~Bondi, M.~G.~J. van~der Burg, and A.~W.~K. Metzner.
\newblock Gravitational waves in general relativity. vii. waves from
  axi-symmetric isolated systems.
\newblock {\em Proceedings of the Royal Society of London A: Mathematical,
  Physical and Engineering Sciences}, 269(1336):21--52,
  1962http://rspa.royalsocietypublishing.org/content/269/1336/21.full.pdf.

\bibitem{Briscese:2006xu}
F.~Briscese, E.~Elizalde, S.~Nojiri, and S.~D. Odintsov.
\newblock {Phantom scalar dark energy as modified gravity: Understanding the
  origin of the Big Rip singularity}.
\newblock {\em Phys. Lett.}, B646:105--111, 2007 arXiv:hep-th/0612220,
  [hep-th].

\bibitem{Brodbeck:1998az}
O.~Brodbeck, S.~Frittelli, P.~Hubner, and O.~A. Reula.
\newblock {Einstein's equations with asymptotically stable constraint
  propagation}.
\newblock {\em J. Math. Phys.}, 40:909--923, 1999 arXiv:gr-qc/9809023,
  [gr-qc].

\bibitem{Calabrese:2005ft}
G.~Calabrese, I.~Hinder, and S.~Husa.
\newblock {Numerical stability for finite difference approximations of
  Einstein's equations}.
\newblock {\em J. Comput. Phys.}, 218:607--634, 2006 arXiv:gr-qc/0503056,
  [gr-qc].

\bibitem{PhysRevD.87.104029}
Z.~Cao, P.~Galaviz, and L.-F. Li.
\newblock Binary black hole mergers in $f(r)$ theory.
\newblock {\em Phys. Rev. D}, 87:104029, May 2013.

\bibitem{capozziello2010beyond}
S.~Capozziello and V.~Faraoni.
\newblock {\em Beyond Einstein Gravity: A Survey of Gravitational Theories for
  Cosmology and Astrophysics}.
\newblock Fundamental Theories of Physics. Springer Netherlands, 2010.

\bibitem{Capozziello:2010ut}
S.~Capozziello and S.~Vignolo.
\newblock {The Cauchy problem for metric-affine f(R)-gravity in presence of a
  Klein-Gordon scalar field}.
\newblock {\em Int. J. Geom. Meth. Mod. Phys.}, 8:167--176, 2011
  arXiv:1003.4280,  [gr-qc].

\bibitem{Capozziello:2011gw}
S.~Capozziello and S.~Vignolo.
\newblock {The Cauchy problem for f(R)-gravity: An Overview}.
\newblock {\em Int. J. Geom. Meth. Mod. Phys.}, 9:1250006, 2012
  arXiv:1103.2302,  [gr-qc].

\bibitem{Capozziello:2013gza}
S.~Capozziello, T.~Harko, F.~S.~N. Lobo, G.~J. Olmo, and S.~Vignolo.
\newblock {The Cauchy problem in hybrid metric-Palatini f(X)-gravity}.
\newblock {\em Int. J. Geom. Meth. Mod. Phys.}, 11(5):1450042, 2014
  arXiv:1312.1320,  [gr-qc].

\bibitem{Capozziello:2006dj}
S.~Capozziello, S.~Nojiri, S.~D. Odintsov, and A.~Troisi.
\newblock {Cosmological viability of f(R)-gravity as an ideal fluid and its
  compatibility with a matter dominated phase}.
\newblock {\em Phys. Lett.}, B639:135--143, 2006 arXiv:astro-ph/0604431,
  [astro-ph].

\bibitem{lrr-2015-1}
V.~Cardoso, L.~Gualtieri, C.~A.~R. Herdeiro, and U.~Sperhake.
\newblock Exploring new physics frontiers through numerical relativity.
\newblock {\em Living Reviews in Relativity}, 18(1), 2015.

\bibitem{Carroll:2004de}
S.~M. Carroll, A.~De~Felice, V.~Duvvuri, D.~A. Easson, M.~Trodden, and M.~S.
  Turner.
\newblock {The Cosmology of generalized modified gravity models}.
\newblock {\em Phys. Rev.}, D71:063513, 2005 arXiv:astro-ph/0410031,
  [astro-ph].

\bibitem{Clifton:2012ry}
T.~Clifton, P.~Dunsby, R.~Goswami, and A.~M. Nzioki.
\newblock {On the absence of the usual weak-field limit, and the impossibility
  of embedding some known solutions for isolated masses in cosmologies with
  f(R) dark energy}.
\newblock {\em Phys. Rev.}, D87(6):063517, 2013 arXiv:1210.0730,  [gr-qc].

\bibitem{Cognola:2006eg}
G.~Cognola, E.~Elizalde, S.~Nojiri, S.~D. Odintsov, and S.~Zerbini.
\newblock {Dark energy in modified Gauss-Bonnet gravity: Late-time acceleration
  and the hierarchy problem}.
\newblock {\em Phys. Rev.}, D73:084007, 2006 arXiv:hep-th/0601008,  [hep-th].

\bibitem{lrr-2000-5}
G.~B. Cook.
\newblock Initial data for numerical relativity.
\newblock {\em Living Reviews in Relativity}, 3(5), 2000.

\bibitem{DeFelice:2010aj}
A.~De~Felice and S.~Tsujikawa.
\newblock {f(R) theories}.
\newblock {\em Living Rev. Rel.}, 13:3, 2010 arXiv:1002.4928,  [gr-qc].

\bibitem{PhysRevD.35.1095}
S.~Detweiler.
\newblock Evolution of the constraint equations in general relativity.
\newblock {\em Phys. Rev. D}, 35:1095--1099, Feb 1987.

\bibitem{lrr-2004-1}
J.~Frauendiener.
\newblock Conformal infinity.
\newblock {\em Living Reviews in Relativity}, 7(1), 2004.

\bibitem{hyperboloidal}
H.~Friedrich.
\newblock Cauchy problems for the conformal vacuum field equations in general
  relativity.
\newblock {\em Communications in Mathematical Physics}, 91(4):445--472, 1983.

\bibitem{Frittelli:1996nj}
S.~Frittelli.
\newblock {Note on the propagation of the constraints in standard (3+1) general
  relativity}.
\newblock {\em Phys. Rev.}, D55:5992--5996, 1997.

\bibitem{Ganguly:2013taa}
A.~Ganguly, R.~Gannouji, R.~Goswami, and S.~Ray.
\newblock {Neutron stars in the Starobinsky model}.
\newblock {\em Phys. Rev.}, D89(6):064019, 2014 arXiv:1309.3279,  [gr-qc].

\bibitem{Goheer:2009qh}
N.~Goheer, R.~Goswami, P.~K.~S. Dunsby, and K.~Ananda.
\newblock {On the co-existence of matter dominated and accelerating solutions
  in f(G)-gravity}.
\newblock {\em Phys. Rev.}, D79:121301, 2009 arXiv:0904.2559,  [gr-qc].

\bibitem{gourgoulhon20123+1}
{\'E}.~Gourgoulhon.
\newblock {\em 3+1 Formalism in General Relativity: Bases of Numerical
  Relativity}.
\newblock Lecture Notes in Physics. Springer Berlin Heidelberg, 2012.

\bibitem{PhysRevD.70.044032}
C.~Gundlach and J.~M. Mart\'{\i}n-Garc\'{\i}a.
\newblock Symmetric hyperbolicity and consistent boundary conditions for
  second-order einstein equations.
\newblock {\em Phys. Rev. D}, 70:044032, Aug 2004.

\bibitem{Gundlach:2005ta}
C.~Gundlach and J.~M. Martin-Garcia.
\newblock {Hyperbolicity of second-order in space systems of evolution
  equations}.
\newblock {\em Class. Quant. Grav.}, 23:S387--S404, 2006 arXiv:gr-qc/0506037,
  [gr-qc].

\bibitem{PhysRevD.90.024017}
J.-Q. Guo, D.~Wang, and A.~V. Frolov.
\newblock Spherical collapse in $f(r)$ gravity and the
  belinskii-khalatnikov-lifshitz conjecture.
\newblock {\em Phys. Rev. D}, 90:024017, Jul 2014.

\bibitem{gustafsson1995time}
B.~Gustafsson, H.~Kreiss, and J.~Oliger.
\newblock {\em Time Dependent Problems and Difference Methods}.
\newblock A Wiley-Interscience Publication. Wiley, 1995.

\bibitem{PhysRevD.53.5583}
A.~Hindawi, B.~A. Ovrut, and D.~Waldram.
\newblock Consistent spin-two coupling and quadratic gravitation.
\newblock {\em Phys. Rev. D}, 53:5583--5596, May 1996.

\bibitem{Hindawi:1995cu}
A.~Hindawi, B.~A. Ovrut, and D.~Waldram.
\newblock {Nontrivial vacua in higher derivative gravitation}.
\newblock {\em Phys. Rev.}, D53:5597--5608, 1996 arXiv:hep-th/9509147,
  [hep-th].

\bibitem{Jaime:2010kn}
L.~G. Jaime, L.~Patino, and M.~Salgado.
\newblock {Robust approach to f(R) gravity}.
\newblock {\em Phys. Rev.}, D83:024039, 2011 arXiv:1006.5747,  [gr-qc].

\bibitem{0004-637X-779-1-39}
B.~Jain, V.~Vikram, and J.~Sakstein.
\newblock Astrophysical tests of modified gravity: Constraints from distance
  indicators in the nearby universe.
\newblock {\em The Astrophysical Journal}, 779(1):39, 2013.

\bibitem{0264-9381-25-9-093001}
J.~L. Jaramillo, J.~A.~V. Kroon, and E.~Gourgoulhon.
\newblock From geometry to numerics: interdisciplinary aspects in mathematical
  and numerical relativity.
\newblock {\em Classical and Quantum Gravity}, 25(9):093001, 2008.

\bibitem{Kidder:2001tz}
L.~E. Kidder, M.~A. Scheel, and S.~A. Teukolsky.
\newblock {Extending the lifetime of 3-D black hole computations with a new
  hyperbolic system of evolution equations}.
\newblock {\em Phys. Rev.}, D64:064017, 2001 arXiv:gr-qc/0105031,  [gr-qc].

\bibitem{Koivisto:2005yk}
T.~Koivisto.
\newblock {Covariant conservation of energy momentum in modified gravities}.
\newblock {\em Class. Quant. Grav.}, 23:4289--4296, 2006 arXiv:gr-qc/0505128,
  [gr-qc].

\bibitem{Laguna:2002zc}
P.~Laguna and D.~Shoemaker.
\newblock {Numerical stability of a new conformal traceless 3+1 formulation of
  the Einstein equation}.
\newblock {\em Class. Quant. Grav.}, 19:3679--3686, 2002 arXiv:gr-qc/0202105,
  [gr-qc].

\bibitem{0264-9381-24-22-024}
N.~Lanahan-Tremblay and V.~Faraoni.
\newblock The cauchy problem of f ( r ) gravity.
\newblock {\em Classical and Quantum Gravity}, 24(22):5667, 2007.

\bibitem{Lehner:2001wq}
L.~Lehner.
\newblock {Numerical relativity: A Review}.
\newblock {\em Class. Quant. Grav.}, 18:R25--R86, 2001 arXiv:gr-qc/0106072,
  [gr-qc].

\bibitem{PhysRevLett.102.221101}
V.~Miranda, S.~E. Jor\'as, I.~Waga, and M.~Quartin.
\newblock Viable singularity-free $f(r)$ gravity without a cosmological
  constant.
\newblock {\em Phys. Rev. Lett.}, 102:221101, Jun 2009.

\bibitem{bish_thesis:2014}
B.~{Mongwane}.
\newblock {\em {Problems in Cosmology and Numerical Relativity}}.
\newblock PhD thesis, University of Cape Town, 2014.

\bibitem{mongwane:2015}
B.~Mongwane.
\newblock Toward a consistent framework for high order mesh refinement schemes
  in numerical relativity.
\newblock {\em General Relativity and Gravitation}, 47(5), 2015.

\bibitem{Myrzakulov:2013hca}
R.~Myrzakulov, L.~Sebastiani, and S.~Zerbini.
\newblock {Some aspects of generalized modified gravity models}.
\newblock {\em Int. J. Mod. Phys.}, D22:1330017, 2013 arXiv:1302.4646,
  [gr-qc].

\bibitem{Nagy:2004td}
G.~Nagy, O.~E. Ortiz, and O.~A. Reula.
\newblock {Strongly hyperbolic second order Einstein's evolution equations}.
\newblock {\em Phys. Rev.}, D70:044012, 2004 arXiv:gr-qc/0402123,  [gr-qc].

\bibitem{Nakamura01011987}
T.~Nakamura, K.~Oohara, and Y.~Kojima.
\newblock General relativistic collapse to black holes and gravitational waves
  from black holes.
\newblock {\em Progress of Theoretical Physics Supplement}, 90:1--218,
  1987http://ptps.oxfordjournals.org/content/90/1.full.pdf+html.

\bibitem{noakes:1983}
D.~R. Noakes.
\newblock The initial value formulation of higher derivative gravity.
\newblock {\em Journal of Mathematical Physics}, 24(7), 1983.

\bibitem{Nojiri:2005jg}
S.~Nojiri and S.~D. Odintsov.
\newblock {Modified Gauss-Bonnet theory as gravitational alternative for dark
  energy}.
\newblock {\em Phys. Lett.}, B631:1--6, 2005 arXiv:hep-th/0508049,  [hep-th].

\bibitem{Nzioki:2013lca}
A.~M. Nzioki, R.~Goswami, and P.~K.~S. Dunsby.
\newblock {Jebsen-Birkhoff theorem and its stability in f(R) gravity}.
\newblock {\em Phys. Rev.}, D89(6):064050, 2014 arXiv:1312.6790,  [gr-qc].

\bibitem{2011CQGra..28h5006P}
V.~{Paschalidis}, S.~M.~H. {Halataei}, S.~L. {Shapiro}, and I.~{Sawicki}.
\newblock {Constraint propagation equations of the 3+1 decomposition of f(R)
  gravity}.
\newblock {\em Classical and Quantum Gravity}, 28(8):085006, April 2011
  arXiv:1103.0984,  [gr-qc].

\bibitem{Resco:2016upv}
M.~A. Resco, A.~de~la Cruz-Dombriz, F.~J.~L. Estrada, and V.~Z. Castrillo.
\newblock {On neutron stars in f(R) theories: small radii, large masses and
  large energy available for emission in a merger}.
\newblock 2016 arXiv:1602.03880,  [gr-qc].

\bibitem{lrr-1998-3}
O.~A. Reula.
\newblock Hyperbolic methods for einstein's equations.
\newblock {\em Living Reviews in Relativity}, 1(3), 1998.

\bibitem{Ruiz:2010qj}
M.~Ruiz, D.~Hilditch, and S.~Bernuzzi.
\newblock {Constraint preserving boundary conditions for the Z4c formulation of
  general relativity}.
\newblock {\em Phys. Rev.}, D83:024025, 2011 arXiv:1010.0523,  [gr-qc].

\bibitem{Sachs103}
R.~K. Sachs.
\newblock Gravitational waves in general relativity. viii. waves in
  asymptotically flat space-time.
\newblock {\em Proceedings of the Royal Society of London A: Mathematical,
  Physical and Engineering Sciences}, 270(1340):103--126,
  1962http://rspa.royalsocietypublishing.org/content/270/1340/103.full.pdf.

\bibitem{0264-9381-23-14-010}
M.~Salgado.
\newblock The cauchy problem of scalar¡vtensor theories of gravity.
\newblock {\em Classical and Quantum Gravity}, 23(14):4719, 2006.

\bibitem{Salgado:2008xh}
M.~Salgado, D.~M.-d. Rio, M.~Alcubierre, and D.~Nunez.
\newblock {Hyperbolicity of scalar-tensor theories of gravity}.
\newblock {\em Phys. Rev.}, D77:104010, 2008 arXiv:0801.2372,  [gr-qc].

\bibitem{PhysRevD.66.064002}
O.~Sarbach, G.~Calabrese, J.~Pullin, and M.~Tiglio.
\newblock Hyperbolicity of the baumgarte-shapiro-shibata-nakamura system of
  einstein evolution equations.
\newblock {\em Phys. Rev. D}, 66:064002, Sep 2002.

\bibitem{Seiberg:2006wf}
N.~Seiberg.
\newblock {Emergent spacetime}.
\newblock  In {\em {The Quantum Structure of Space and Time}}, pages 163--178,
  2006 arXiv:hep-th/0601234,  [hep-th].

\bibitem{Shibata:1995we}
M.~Shibata and T.~Nakamura.
\newblock {Evolution of three-dimensional gravitational waves: Harmonic slicing
  case}.
\newblock {\em Phys.Rev.}, D52:5428--5444, 1995.

\bibitem{Shinkai:2001nd}
H.-a. Shinkai and G.~Yoneda.
\newblock {Adjusted ADM systems and their expected stability properties}.
\newblock {\em Class. Quant. Grav.}, 19:1027--1050, 2002 arXiv:gr-qc/0110008,
  [gr-qc].

\bibitem{Shinkai:2002yf}
H.-a. Shinkai and G.~Yoneda.
\newblock {Reformulating the Einstein equations for stable numerical
  simulations}.
\newblock 2002 arXiv:gr-qc/0209111,  [gr-qc].

\bibitem{shinkai:2004}
H.-a. Shinkai and G.~Yoneda.
\newblock Letter: Constraint propagation in (n + 1)-dimensional space-time.
\newblock {\em General Relativity and Gravitation}, 36(8):1931--1937, 2004.

\bibitem{Siebel:2001dr}
F.~Siebel and P.~Hubner.
\newblock {On the effect of constraint enforcement on the quality of numerical
  solutions in general relativity}.
\newblock {\em Phys. Rev.}, D64:024021, 2001 arXiv:gr-qc/0104079,  [gr-qc].

\bibitem{Sotiriou:2006hs}
T.~P. Sotiriou.
\newblock {f(R) gravity and scalar-tensor theory}.
\newblock {\em Class. Quant. Grav.}, 23:5117--5128, 2006 arXiv:gr-qc/0604028,
  [gr-qc].

\bibitem{Sotiriou:2008rp}
T.~P. Sotiriou and V.~Faraoni.
\newblock {f(R) Theories Of Gravity}.
\newblock {\em Rev. Mod. Phys.}, 82:451--497, 2010 arXiv:0805.1726,  [gr-qc].

\bibitem{STAROBINSKY198099}
A.~Starobinsky.
\newblock A new type of isotropic cosmological models without singularity.
\newblock {\em Physics Letters B}, 91(1):99 -- 102, 1980.

\bibitem{Stelle1978}
K.~S. Stelle.
\newblock Classical gravity with higher derivatives.
\newblock {\em General Relativity and Gravitation}, 9(4):353--371, 1978.

\bibitem{Tsokaros:2013fma}
A.~Tsokaros.
\newblock {On the initial value problem of $f(R)$ theories and their degrees of
  freedom}.
\newblock {\em Class. Quant. Grav.}, 31:025021, 2014.

\bibitem{wald:1984}
R.~M. {Wald}.
\newblock {\em {General relativity}}.
\newblock Chicago, University of Chicago Press, 1984, 504 p., 1984.

\bibitem{Woodard:2006nt}
R.~P. Woodard.
\newblock {Avoiding dark energy with 1/r modifications of gravity}.
\newblock {\em Lect. Notes Phys.}, 720:403--433, 2007 arXiv:astro-ph/0601672,
  [astro-ph].

\bibitem{Yoneda:2001iy}
G.~Yoneda and H.-a. Shinkai.
\newblock {Constraint propagation in the family of ADM systems}.
\newblock {\em Phys. Rev.}, D63:124019, 2001 arXiv:gr-qc/0103032,  [gr-qc].

\bibitem{Yoneda:2000zr}
G.~Yoneda and H.-a. Shinkai.
\newblock {Hyperbolic formulations and numerical relativity 2: Asymptotically
  constrained system of the Einstein equation}.
\newblock {\em Class. Quant. Grav.}, 18:441--462, 2001 arXiv:gr-qc/0007034,
  [gr-qc].

\bibitem{Yoneda:2002kg}
G.~Yoneda and H.-a. Shinkai.
\newblock {Advantages of modified ADM formulation: Constraint propagation
  analysis of Baumgarte-Shapiro-Shibata-Nakamura system}.
\newblock {\em Phys. Rev.}, D66:124003, 2002 arXiv:gr-qc/0204002,  [gr-qc].

\bibitem{1979sgrr.work...83Y}
J.~W. {York}, Jr.
\newblock {Kinematics and dynamics of general relativity} In {\em Sources of
  Gravitational Radiation}.
\newblock L.~L. {Smarr}, editor, pages 83--126, 1979.

\end{thebibliography}

\end{document}